\documentstyle[]{article}
\begin{document}
\overfullrule 0 mm
\language 0
\vskip 0.5 cm
\centerline { \bf{RUTHERFORD SCATTERING }} \centerline {
\bf {WITH RETARDATION}}
 \vskip 0.5 cm

\centerline {\bf{ Alexander A.  Vlasov}}
\vskip 0.3 cm
\centerline {{  High Energy and Quantum Theory}}
\centerline {{Department of Physics}}
\centerline {{ Moscow State University}}
\centerline {{  Moscow, 119899}}
\centerline {{ Russia}}
\vskip 0.3cm
{\it  Numerical solutions for  Sommerfeld model in nonrelativistic
case are presented for the scattering of a spinless extended
charged body in a static Coulomb field of a fixed point charge.
It is shown that differential cross section for extended body
preserves the form of the  Rutherford result with multiplier, not
equal to one (as in classical case), but depending on the size of Sommerfeld 
particle. Also the effect of capture by attractive center is found out for 
Sommerfeld particle. The origin of this effect lies in radiation damping.}

03.50.De
\vskip 0.3 cm

Here we continue [1] our numerical investigation of Sommerfeld model
in classical electrodynamics. Let us remind that
Sommerfeld model of charged rigid sphere [2] is the simplest model to
take into consideration the "back-reaction" of self-electromagnetic field
of a radiating extended charged body on its equation of motion (in
the limit of zero body's size we have the known Lorentz-Dirac
equation with all its problems: renormalization of
mass, preacceleration, run-away solutions, etc.).

In the previous article  the effect of classical tunneling was
considered - due to retardation moving body begins "to feel" the existence
of potential barrier too late, when this barrier is overcome ([1],
see also [3]).

Consequently one should expect that Rutherford scattering of a charged 
extended body in the static Coulomb field of a fixed point charge also 
differs from classical scattering of point-like particle 
(for Lorentz-Dirac equation Rutherford scattering was numerically 
investigated in [4]).

For the case of simplicity here we consider the nonrelativistic,
linear in velocity, version of Sommerfeld model.

Let the total charge of a  uniformly  charged sphere be $Q$,
mechanical mass - $m$, radius - $a$.  Then its equation of motion
reads:  $$m\dot{\vec v} =\vec F_{ext}+ \eta\left[\vec v(t-2a/c) -
\vec v(t)\right] \eqno(1)$$ here  $\eta=
{Q^2 \over 3  c a^2},\ \ \vec v= d \vec R /dt,\ \ \vec R$ -
coordinate of the center of the shell.

 External force $\vec F_{ext}$, produced by fixed point charge $e$
(placed at $\vec r=0$), is $$ \vec F_{ext} =\int d \vec r \rho \cdot
{e\vec r \over r^3}$$ and for $$\rho = Q\delta( |\vec r - \vec R|
-a)/4\pi a^2 $$ reads $$\vec F_{ext}=  {e\vec R \over R^3},\ \ R>a
\eqno(2)$$

In dimensionless variables $\vec R= \vec \Pi\cdot 2L,\ \
ct=x \cdot 2L$  equation (1-2) takes the form
$$\ddot{ \vec \Pi} =K\left[ \dot{ \vec \Pi}(x-\delta)-\dot {\vec \Pi}(x)
\right] +\lambda \cdot \vec \Pi \cdot |\vec \Pi|^{-3}  \eqno (3)$$ with $$
K={2Q^2L\over 3mc^2a^2},\ \ \lambda={eQ\over 2mc^2L},\ \ \delta= {a/L}$$
or
$$K={2 r_{cl} L\over 3 a^2},\ \ \lambda={e r_{cl}\over Q2L},\ \
r_{cl}={Q^2\over mc^2}$$
Taking the $X-Y$ plane to be the plane of
scattering ($\vec \Pi= (X,Y,0)$ ), we split equation (3)  into two:
$$\ddot{  Y} =K\left[ \dot{  Y}(x-\delta)-\dot {Y}(x) \right]
+\lambda \cdot Y \cdot (X^2+Y^2)^{-3/2}  $$ $$\ddot{  X} =K\left[
\dot{  X}(x-\delta)-\dot {X}(x) \right] +\lambda \cdot X \cdot
(X^2+Y^2)^{-3/2}  \eqno(4)$$
 The starting conditions at $x=0$ are:
$$X_i=1000,\ \ Y_i=b \ \ (b-impact \ parameter),\ \ 
\dot {X}_i = v_i=-0.1,\ \ \dot {Y}_i =0 \eqno(5)$$

Numerical results are expressed on Figs. 1,2,3.

{\bf 1.}

On Fig. 1. one can see how the scattering angle varies from
point-like particle (classical scattering, curve 1) to  extended body
(curve 2). Here we have chosen
$$L=5r_{cl},\ \  b=60.0,\ \ \delta=4.0,\ \ \lambda=0.1, \ \ 
K=(2/15)(\delta)^{-2}$$
i.e.
$$a=20 r_{cl},\ \ e=Q,\ \ K=(10/3)(r_{cl}/a)^2 =1/120$$
vertical axis is $Y$, 
horizontal - $X$. 

Thus due to retardation the scattering angle $\theta$
for extended body is smaller than that for point-like particle.

{\bf 2.}

Differential cross section $d\sigma$ is given by the formula
$$d\sigma =2\pi \rho(\theta)|{d\rho(\theta) \over d \theta}| d\theta
$$ where $\rho=b\cdot 2L,$
or
$${ 1\over  2\pi (2L)^2} \cdot {d\sigma\over d\xi}= {d
b^2 \over d \xi} \eqno(6)$$ where $$\xi= {1+\cos{\theta}\over
1-\cos{\theta}}$$ Classical Rutherford result is that R.H.S. of eq.
(4)  is constant:  $$ b^2\cdot (v_i)^4\cdot (\lambda)^{-2} = \xi
\eqno(7) $$
or
$${ (\lambda)^2 \over  2\pi (2L)^2 (v_i)^4} \cdot {d\sigma\over
d\xi}=1 \eqno(8)$$
This classical result can be derived from eq.(4) in standard manner for
$K=0$ (see, for ex., [5])

In the case of extended body 
$$L=5r_{cl},\ \ \ \lambda=0.1,\ \  
K=(2/15)(\delta)^{-2}$$
i.e.
$$ e=Q,\ \ K=(10/3)(r_{cl}/a)^2 $$

 numerical calculations for various
values of $b,\ \ 10.0<b<110.0$ show that Rutherford formula (7,8)
changes in the following way:
 $$ b^2\cdot (v_i)^4\cdot (\lambda)^{-2} = \xi\cdot
\left[ 1+ const / \delta \right]^{-1} \eqno(9) $$
or
$${ (\lambda)^2 \over  2\pi (2L)^2 (v_i)^4} \cdot {d\sigma\over
d\xi}=\left[ 1+ const / \delta \right]^{-1} \eqno(10)$$
 where the
 multiplier $const$  is equal approximately to $0.30$.

 Thus differential cross section for extended body preserves the form
 of the  Rutherford result with multiplier, not equal to one
 (as in classical case), but depending on the value of  size of Sommerfeld 
particle. For $\delta \to \infty$ (i.e. $K \to 0$ ) formula (9,10) gives  the 
Rutherford result.

 On Fig. 2 we see how the direct proportionality between  $
 b^2\cdot (v_i)^4\cdot (\lambda)^{-2}$ and $\xi$ changes in
 accordance to formula (9). Vertical axis is    $b^2\cdot
 (v_i)^4\cdot (\lambda)^{-2}$ and horizontal - $\xi$. Values of
 retardation $\delta$ (or dimensionless body's size) are taken to be
 $\ 1,\ 2,\ 3,\ 4$,  and curves
 are marked accordingly as $\ 1,\ 2,\ 3,\ 4$;
in the case of Rutherford scattering ($K \equiv 0$) the curve is marked 
as "R".

{\bf 3.}

On Fig.3 we see the appearance of the effect of capture of Sommerfeld 
particle with charge $Q$ by the attractive Coulomb center with charge $e$. 
Here we have chosen the following values of parameters:
$$ L=r_{cl}/2,\ \ \lambda=-1.0, \ \ K=(4/3)(\delta)^{-2},\ \ \delta=5.0,\ \ 
b=30.0 $$ i.e. $$e=-Q,\ \ a=2.5 r_{cl}, \ \ \ K=4/75.$$  Initial 
conditions are the same as in (5). 

Following classical Rutherford scattering (eq. (4) with $K\equiv 0$) for 
initial conditions (5) trajectory must be infinite one and thus there is 
no capture; but this is not the case of Sommerfeld particle: due to 
radiation damping particle loses its energy and consequently can fall down 
on the attractive center.

Varying impact parameter $b$ with fixed $\lambda=-1.0$ and $\delta=5.0$  we 
numerically found out the crucial value of $b$ when the effect of capture 
begins:
$$b \leq b_{cr} \approx 31.40$$

  \vskip 0.5 cm

 \centerline {\bf{REFERENCES}}

  \begin{enumerate}
\item Alexander A.Vlasov, physics/9905050.
\item
 A.Sommerfeld, Gottingen Nachrichten, 29 (1904), 363 (1904), 201
  (1905).

 L.Page, Phys.Rev., 11, 377 (1918)

 T.Erber, Fortschr. Phys., 9, 343 (1961)

 P.Pearle in "Electromagnetism",ed. D.Tepliz, (Plenum, N.Y.,
1982), p.211.

 A.Yaghjian, "Relativistic Dynamics of a Charged Sphere".
  Lecture Notes in Physics, 11 (Springer-Verlag, Berlin, 1992).

 F.Rohrlich, Am.J.Phys., 65(11), 1051(1997).
\item Alexander A.Vlasov, physics/9905050.

 F.Denef et al, Phys.Rev. E56, 3624 (1997); hep-th/9602066.

 Alexander A.Vlasov, Theoretical and Mathematical Physics, 109, n.3,
 1608(1996).
\item J.Huschielt and W.E.Baylis, Phys.Rev. D17, N 4,
985 (1978).
\item L.D. Landau, E.M.Lifshitz, Mechanics, 4-th edition, Moscow, Nauka, 
1988.
  \end{enumerate} 
\eject

\newcount\numpoint                     
\newcount\numpointo                    
\numpoint=1 \numpointo=1               
\def\emmoveto#1#2{\offinterlineskip   
\hbox to 0 true cm{\vbox to 0          
true cm{\vskip - #2 true mm             
\hskip #1 true mm \special{em:point    
\the\numpoint}\vss}\hss}             
\numpointo=\numpoint                   
\global\advance \numpoint by 1}       
\def\emlineto#1#2{\offinterlineskip   
\hbox to 0 true cm{\vbox to 0          
true cm{\vskip - #2 true mm             
\hskip #1 true mm \special{em:point    
\the\numpoint}\vss}\hss}             
\special{em:line                        
\the\numpointo,\the\numpoint}        
\numpointo=\numpoint                   
\global\advance \numpoint by 1}       
\def\emshow#1#2#3{\offinterlineskip   
\hbox to 0 true cm{\vbox to 0          
true cm{\vskip - #2 true mm             
\hskip #1 true mm \vbox to 0           
true cm{\vss\hbox{#3\hss              
}}\vss}\hss}}                          
\special{em:linewidth 0.8pt}            

\vrule width 0 mm height                0 mm depth 90.000 true mm                

\special{em:linewidth 0.8pt} 
\emmoveto{130.000}{10.000} 
\emlineto{12.000}{10.000} 
\emlineto{12.000}{80.000} 
\emmoveto{71.000}{10.000} 
\emlineto{71.000}{80.000} 
\emmoveto{12.000}{45.000} 
\emlineto{130.000}{45.000} 
\emmoveto{130.000}{10.000} 
\emlineto{130.000}{80.000} 
\emlineto{12.000}{80.000} 
\emlineto{12.000}{10.000} 
\emlineto{130.000}{10.000} 
\special{em:linewidth 0.4pt} 
\emmoveto{12.000}{17.000} 
\emlineto{130.000}{17.000} 
\emmoveto{12.000}{24.000} 
\emlineto{130.000}{24.000} 
\emmoveto{12.000}{31.000} 
\emlineto{130.000}{31.000} 
\emmoveto{12.000}{38.000} 
\emlineto{130.000}{38.000} 
\emmoveto{12.000}{45.000} 
\emlineto{130.000}{45.000} 
\emmoveto{12.000}{52.000} 
\emlineto{130.000}{52.000} 
\emmoveto{12.000}{59.000} 
\emlineto{130.000}{59.000} 
\emmoveto{12.000}{66.000} 
\emlineto{130.000}{66.000} 
\emmoveto{12.000}{73.000} 
\emlineto{130.000}{73.000} 
\emmoveto{23.800}{10.000} 
\emlineto{23.800}{80.000} 
\emmoveto{35.600}{10.000} 
\emlineto{35.600}{80.000} 
\emmoveto{47.400}{10.000} 
\emlineto{47.400}{80.000} 
\emmoveto{59.200}{10.000} 
\emlineto{59.200}{80.000} 
\emmoveto{71.000}{10.000} 
\emlineto{71.000}{80.000} 
\emmoveto{82.800}{10.000} 
\emlineto{82.800}{80.000} 
\emmoveto{94.600}{10.000} 
\emlineto{94.600}{80.000} 
\emmoveto{106.400}{10.000} 
\emlineto{106.400}{80.000} 
\emmoveto{118.200}{10.000} 
\emlineto{118.200}{80.000} 
\special{em:linewidth 0.8pt} 
\emmoveto{129.676}{22.085} 
\emlineto{129.356}{22.095} 
\emmoveto{129.356}{22.085} 
\emlineto{129.036}{22.095} 
\emmoveto{129.036}{22.085} 
\emlineto{128.716}{22.095} 
\emmoveto{128.716}{22.085} 
\emlineto{128.396}{22.095} 
\emmoveto{128.396}{22.085} 
\emlineto{128.076}{22.095} 
\emmoveto{128.076}{22.085} 
\emlineto{127.756}{22.095} 
\emmoveto{127.756}{22.085} 
\emlineto{127.436}{22.095} 
\emmoveto{127.436}{22.085} 
\emlineto{127.117}{22.095} 
\emmoveto{127.117}{22.085} 
\emlineto{126.797}{22.095} 
\emmoveto{126.797}{22.085} 
\emlineto{126.477}{22.095} 
\emmoveto{126.477}{22.085} 
\emlineto{126.157}{22.095} 
\emmoveto{126.157}{22.085} 
\emlineto{125.837}{22.095} 
\emmoveto{125.837}{22.085} 
\emlineto{125.517}{22.095} 
\emmoveto{125.517}{22.085} 
\emlineto{125.198}{22.095} 
\emmoveto{125.198}{22.085} 
\emlineto{124.878}{22.096} 
\emmoveto{124.878}{22.086} 
\emlineto{124.558}{22.096} 
\emmoveto{124.558}{22.086} 
\emlineto{124.238}{22.096} 
\emmoveto{124.238}{22.086} 
\emlineto{123.919}{22.096} 
\emmoveto{123.919}{22.086} 
\emlineto{123.599}{22.096} 
\emmoveto{123.599}{22.086} 
\emlineto{123.279}{22.096} 
\emmoveto{123.279}{22.086} 
\emlineto{122.960}{22.096} 
\emmoveto{122.960}{22.086} 
\emlineto{122.640}{22.096} 
\emmoveto{122.640}{22.086} 
\emlineto{122.320}{22.096} 
\emmoveto{122.320}{22.086} 
\emlineto{122.001}{22.096} 
\emmoveto{122.001}{22.086} 
\emlineto{121.681}{22.096} 
\emmoveto{121.681}{22.086} 
\emlineto{121.362}{22.096} 
\emmoveto{121.362}{22.086} 
\emlineto{121.042}{22.096} 
\emmoveto{121.042}{22.086} 
\emlineto{120.723}{22.097} 
\emmoveto{120.723}{22.087} 
\emlineto{120.403}{22.097} 
\emmoveto{120.403}{22.087} 
\emlineto{120.084}{22.097} 
\emmoveto{120.084}{22.087} 
\emlineto{119.764}{22.097} 
\emmoveto{119.764}{22.087} 
\emlineto{119.445}{22.097} 
\emmoveto{119.445}{22.087} 
\emlineto{119.125}{22.097} 
\emmoveto{119.125}{22.087} 
\emlineto{118.806}{22.097} 
\emmoveto{118.806}{22.087} 
\emlineto{118.487}{22.098} 
\emmoveto{118.487}{22.088} 
\emlineto{118.167}{22.098} 
\emmoveto{118.167}{22.088} 
\emlineto{117.848}{22.098} 
\emmoveto{117.848}{22.088} 
\emlineto{117.529}{22.098} 
\emmoveto{117.529}{22.088} 
\emlineto{117.210}{22.098} 
\emmoveto{117.210}{22.088} 
\emlineto{116.890}{22.098} 
\emmoveto{116.890}{22.088} 
\emlineto{116.571}{22.099} 
\emmoveto{116.571}{22.089} 
\emlineto{116.252}{22.099} 
\emmoveto{116.252}{22.089} 
\emlineto{115.933}{22.099} 
\emmoveto{115.933}{22.089} 
\emlineto{115.614}{22.099} 
\emmoveto{115.614}{22.089} 
\emlineto{115.294}{22.099} 
\emmoveto{115.294}{22.089} 
\emlineto{114.975}{22.099} 
\emmoveto{114.975}{22.089} 
\emlineto{114.656}{22.100} 
\emmoveto{114.656}{22.090} 
\emlineto{114.337}{22.100} 
\emmoveto{114.337}{22.090} 
\emlineto{114.018}{22.100} 
\emmoveto{114.018}{22.090} 
\emlineto{113.699}{22.100} 
\emmoveto{113.699}{22.090} 
\emlineto{113.380}{22.101} 
\emmoveto{113.380}{22.091} 
\emlineto{113.061}{22.101} 
\emmoveto{113.061}{22.091} 
\emlineto{112.742}{22.101} 
\emmoveto{112.742}{22.091} 
\emlineto{112.424}{22.101} 
\emmoveto{112.424}{22.091} 
\emlineto{112.105}{22.102} 
\emmoveto{112.105}{22.092} 
\emlineto{111.786}{22.102} 
\emmoveto{111.786}{22.092} 
\emlineto{111.467}{22.102} 
\emmoveto{111.467}{22.092} 
\emlineto{111.148}{22.102} 
\emmoveto{111.148}{22.092} 
\emlineto{110.830}{22.103} 
\emmoveto{110.830}{22.093} 
\emlineto{110.511}{22.103} 
\emmoveto{110.511}{22.093} 
\emlineto{110.192}{22.103} 
\emmoveto{110.192}{22.093} 
\emlineto{109.874}{22.104} 
\emmoveto{109.874}{22.094} 
\emlineto{109.555}{22.104} 
\emmoveto{109.555}{22.094} 
\emlineto{109.237}{22.104} 
\emmoveto{109.237}{22.094} 
\emlineto{108.918}{22.105} 
\emmoveto{108.918}{22.095} 
\emlineto{108.600}{22.105} 
\emmoveto{108.600}{22.095} 
\emlineto{108.281}{22.105} 
\emmoveto{108.281}{22.095} 
\emlineto{107.963}{22.106} 
\emmoveto{107.963}{22.096} 
\emlineto{107.644}{22.106} 
\emmoveto{107.644}{22.096} 
\emlineto{107.326}{22.107} 
\emmoveto{107.326}{22.097} 
\emlineto{107.008}{22.107} 
\emmoveto{107.008}{22.097} 
\emlineto{106.689}{22.107} 
\emmoveto{106.689}{22.097} 
\emlineto{106.371}{22.108} 
\emmoveto{106.371}{22.098} 
\emlineto{106.053}{22.108} 
\emmoveto{106.053}{22.098} 
\emlineto{105.735}{22.109} 
\emmoveto{105.735}{22.099} 
\emlineto{105.417}{22.109} 
\emmoveto{105.417}{22.099} 
\emlineto{105.099}{22.110} 
\emmoveto{105.099}{22.100} 
\emlineto{104.780}{22.110} 
\emmoveto{104.780}{22.100} 
\emlineto{104.462}{22.111} 
\emmoveto{104.462}{22.101} 
\emlineto{104.145}{22.111} 
\emmoveto{104.145}{22.101} 
\emlineto{103.827}{22.112} 
\emmoveto{103.827}{22.102} 
\emlineto{103.509}{22.112} 
\emmoveto{103.509}{22.102} 
\emlineto{103.191}{22.113} 
\emmoveto{103.191}{22.103} 
\emlineto{102.873}{22.113} 
\emmoveto{102.873}{22.103} 
\emlineto{102.555}{22.114} 
\emmoveto{102.555}{22.104} 
\emlineto{102.238}{22.115} 
\emmoveto{102.238}{22.105} 
\emlineto{101.920}{22.115} 
\emmoveto{101.920}{22.105} 
\emlineto{101.602}{22.116} 
\emmoveto{101.602}{22.106} 
\emlineto{101.285}{22.116} 
\emmoveto{101.285}{22.106} 
\emlineto{100.967}{22.117} 
\emmoveto{100.967}{22.107} 
\emlineto{100.650}{22.118} 
\emmoveto{100.650}{22.108} 
\emlineto{100.333}{22.119} 
\emmoveto{100.333}{22.109} 
\emlineto{100.015}{22.119} 
\emmoveto{100.015}{22.109} 
\emlineto{99.698}{22.120} 
\emmoveto{99.698}{22.110} 
\emlineto{99.381}{22.121} 
\emmoveto{99.381}{22.111} 
\emlineto{99.064}{22.122} 
\emmoveto{99.064}{22.112} 
\emlineto{98.746}{22.122} 
\emmoveto{98.746}{22.112} 
\emlineto{98.429}{22.123} 
\emmoveto{98.429}{22.113} 
\emlineto{98.112}{22.124} 
\emmoveto{98.112}{22.114} 
\emlineto{97.795}{22.125} 
\emmoveto{97.795}{22.115} 
\emlineto{97.479}{22.126} 
\emmoveto{97.479}{22.116} 
\emlineto{97.162}{22.127} 
\emmoveto{97.162}{22.117} 
\emlineto{96.845}{22.128} 
\emmoveto{96.845}{22.118} 
\emlineto{96.528}{22.129} 
\emmoveto{96.528}{22.119} 
\emlineto{96.212}{22.130} 
\emmoveto{96.212}{22.120} 
\emlineto{95.895}{22.131} 
\emmoveto{95.895}{22.121} 
\emlineto{95.579}{22.132} 
\emmoveto{95.579}{22.122} 
\emlineto{95.262}{22.133} 
\emmoveto{95.262}{22.123} 
\emlineto{94.946}{22.134} 
\emmoveto{94.946}{22.124} 
\emlineto{94.630}{22.135} 
\emmoveto{94.630}{22.125} 
\emlineto{94.313}{22.136} 
\emmoveto{94.313}{22.126} 
\emlineto{93.997}{22.138} 
\emmoveto{93.997}{22.128} 
\emlineto{93.681}{22.139} 
\emmoveto{93.681}{22.129} 
\emlineto{93.365}{22.140} 
\emmoveto{93.365}{22.130} 
\emlineto{93.049}{22.141} 
\emmoveto{93.049}{22.131} 
\emlineto{92.734}{22.143} 
\emmoveto{92.734}{22.133} 
\emlineto{92.418}{22.144} 
\emmoveto{92.418}{22.134} 
\emlineto{92.102}{22.145} 
\emmoveto{92.102}{22.135} 
\emlineto{91.787}{22.147} 
\emmoveto{91.787}{22.137} 
\emlineto{91.471}{22.148} 
\emmoveto{91.471}{22.138} 
\emlineto{91.156}{22.150} 
\emmoveto{91.156}{22.140} 
\emlineto{90.841}{22.152} 
\emmoveto{90.841}{22.142} 
\emlineto{90.526}{22.153} 
\emmoveto{90.526}{22.143} 
\emlineto{90.211}{22.155} 
\emmoveto{90.211}{22.145} 
\emlineto{89.896}{22.157} 
\emmoveto{89.896}{22.147} 
\emlineto{89.581}{22.159} 
\emmoveto{89.581}{22.149} 
\emlineto{89.266}{22.161} 
\emmoveto{89.266}{22.151} 
\emlineto{88.952}{22.162} 
\emmoveto{88.952}{22.152} 
\emlineto{88.637}{22.164} 
\emmoveto{88.637}{22.154} 
\emlineto{88.323}{22.166} 
\emmoveto{88.323}{22.156} 
\emlineto{88.008}{22.169} 
\emmoveto{88.008}{22.159} 
\emlineto{87.694}{22.171} 
\emmoveto{87.694}{22.161} 
\emlineto{87.380}{22.173} 
\emmoveto{87.380}{22.163} 
\emlineto{87.066}{22.175} 
\emmoveto{87.066}{22.165} 
\emlineto{86.753}{22.178} 
\emmoveto{86.753}{22.168} 
\emlineto{86.439}{22.180} 
\emmoveto{86.439}{22.170} 
\emlineto{86.126}{22.183} 
\emmoveto{86.126}{22.173} 
\emlineto{85.812}{22.186} 
\emmoveto{85.812}{22.176} 
\emlineto{85.499}{22.188} 
\emmoveto{85.499}{22.178} 
\emlineto{85.186}{22.191} 
\emmoveto{85.186}{22.181} 
\emlineto{84.873}{22.194} 
\emmoveto{84.873}{22.184} 
\emlineto{84.561}{22.197} 
\emmoveto{84.561}{22.187} 
\emlineto{84.248}{22.200} 
\emmoveto{84.248}{22.190} 
\emlineto{83.936}{22.204} 
\emmoveto{83.936}{22.194} 
\emlineto{83.624}{22.207} 
\emmoveto{83.624}{22.197} 
\emlineto{83.312}{22.211} 
\emmoveto{83.312}{22.201} 
\emlineto{83.000}{22.214} 
\emmoveto{83.000}{22.204} 
\emlineto{82.689}{22.218} 
\emmoveto{82.689}{22.208} 
\emlineto{82.377}{22.222} 
\emmoveto{82.377}{22.212} 
\emlineto{82.066}{22.226} 
\emmoveto{82.066}{22.216} 
\emlineto{81.755}{22.231} 
\emmoveto{81.755}{22.221} 
\emlineto{81.445}{22.235} 
\emmoveto{81.445}{22.225} 
\emlineto{81.134}{22.240} 
\emmoveto{81.134}{22.230} 
\emlineto{80.824}{22.245} 
\emmoveto{80.824}{22.235} 
\emlineto{80.514}{22.250} 
\emmoveto{80.514}{22.240} 
\emlineto{80.205}{22.255} 
\emmoveto{80.205}{22.245} 
\emlineto{79.895}{22.261} 
\emmoveto{79.895}{22.251} 
\emlineto{79.586}{22.267} 
\emmoveto{79.586}{22.257} 
\emlineto{79.277}{22.273} 
\emmoveto{79.277}{22.263} 
\emlineto{78.969}{22.279} 
\emmoveto{78.969}{22.269} 
\emlineto{78.661}{22.286} 
\emmoveto{78.661}{22.276} 
\emlineto{78.353}{22.293} 
\emmoveto{78.353}{22.283} 
\emlineto{78.046}{22.301} 
\emmoveto{78.046}{22.291} 
\emlineto{77.739}{22.308} 
\emmoveto{77.739}{22.298} 
\emlineto{77.432}{22.317} 
\emmoveto{77.432}{22.307} 
\emlineto{77.126}{22.325} 
\emmoveto{77.126}{22.315} 
\emlineto{76.820}{22.334} 
\emmoveto{76.820}{22.324} 
\emlineto{76.515}{22.344} 
\emmoveto{76.515}{22.334} 
\emlineto{76.210}{22.354} 
\emmoveto{76.210}{22.344} 
\emlineto{75.906}{22.365} 
\emmoveto{75.906}{22.355} 
\emlineto{75.602}{22.376} 
\emmoveto{75.602}{22.366} 
\emlineto{75.298}{22.389} 
\emmoveto{75.298}{22.379} 
\emlineto{74.996}{22.402} 
\emmoveto{74.996}{22.392} 
\emlineto{74.694}{22.415} 
\emmoveto{74.694}{22.405} 
\emlineto{74.392}{22.430} 
\emmoveto{74.392}{22.420} 
\emlineto{74.091}{22.446} 
\emmoveto{74.091}{22.436} 
\emlineto{73.791}{22.462} 
\emmoveto{73.791}{22.452} 
\emlineto{73.492}{22.480} 
\emmoveto{73.492}{22.470} 
\emlineto{73.193}{22.500} 
\emmoveto{73.193}{22.490} 
\emlineto{72.895}{22.520} 
\emmoveto{72.895}{22.510} 
\emlineto{72.599}{22.542} 
\emmoveto{72.599}{22.532} 
\emlineto{72.303}{22.566} 
\emmoveto{72.303}{22.556} 
\emlineto{72.007}{22.592} 
\emmoveto{72.007}{22.582} 
\emlineto{71.713}{22.620} 
\emmoveto{71.713}{22.610} 
\emlineto{71.420}{22.650} 
\emmoveto{71.420}{22.640} 
\emlineto{71.129}{22.683} 
\emmoveto{71.129}{22.673} 
\emlineto{70.838}{22.719} 
\emmoveto{70.838}{22.709} 
\emlineto{70.549}{22.758} 
\emmoveto{70.549}{22.748} 
\emlineto{70.260}{22.801} 
\emmoveto{70.260}{22.791} 
\emlineto{69.974}{22.847} 
\emmoveto{69.974}{22.837} 
\emlineto{69.688}{22.898} 
\emmoveto{69.688}{22.888} 
\emlineto{69.404}{22.954} 
\emmoveto{69.404}{22.944} 
\emlineto{69.121}{23.015} 
\emmoveto{69.121}{23.005} 
\emlineto{68.840}{23.082} 
\emmoveto{68.840}{23.072} 
\emlineto{68.560}{23.156} 
\emmoveto{68.560}{23.146} 
\emlineto{68.282}{23.237} 
\emmoveto{68.282}{23.227} 
\emlineto{68.005}{23.325} 
\emmoveto{68.005}{23.315} 
\emlineto{67.729}{23.422} 
\emmoveto{67.729}{23.412} 
\emlineto{67.454}{23.527} 
\emmoveto{67.454}{23.517} 
\emlineto{67.180}{23.642} 
\emmoveto{67.180}{23.632} 
\emlineto{66.907}{23.766} 
\emmoveto{66.907}{23.756} 
\emlineto{66.635}{23.899} 
\emmoveto{66.635}{23.889} 
\emlineto{66.363}{24.043} 
\emmoveto{66.363}{24.033} 
\emlineto{66.091}{24.197} 
\emmoveto{66.091}{24.187} 
\emlineto{65.820}{24.360} 
\emmoveto{65.820}{24.350} 
\emlineto{65.548}{24.533} 
\emmoveto{65.548}{24.523} 
\emlineto{65.276}{24.715} 
\emmoveto{65.276}{24.705} 
\emlineto{65.004}{24.906} 
\emmoveto{65.004}{24.896} 
\emlineto{64.731}{25.106} 
\emmoveto{64.731}{25.096} 
\emlineto{64.458}{25.313} 
\emmoveto{64.458}{25.303} 
\emlineto{64.183}{25.528} 
\emmoveto{64.183}{25.518} 
\emlineto{63.908}{25.749} 
\emmoveto{63.908}{25.739} 
\emlineto{63.633}{25.977} 
\emmoveto{63.633}{25.967} 
\emlineto{63.356}{26.210} 
\emmoveto{63.356}{26.200} 
\emlineto{63.078}{26.449} 
\emmoveto{63.078}{26.439} 
\emlineto{62.800}{26.693} 
\emmoveto{62.800}{26.683} 
\emlineto{62.520}{26.941} 
\emmoveto{62.520}{26.931} 
\emlineto{62.240}{27.193} 
\emmoveto{62.240}{27.183} 
\emlineto{61.959}{27.449} 
\emmoveto{61.959}{27.439} 
\emlineto{61.677}{27.708} 
\emmoveto{61.677}{27.698} 
\emlineto{61.395}{27.971} 
\emmoveto{61.395}{27.961} 
\emlineto{61.112}{28.236} 
\emmoveto{61.112}{28.226} 
\emlineto{60.828}{28.504} 
\emmoveto{60.828}{28.494} 
\emlineto{60.543}{28.775} 
\emmoveto{60.543}{28.765} 
\emlineto{60.258}{29.048} 
\emmoveto{60.258}{29.038} 
\emlineto{59.972}{29.323} 
\emmoveto{59.972}{29.313} 
\emlineto{59.685}{29.599} 
\emmoveto{59.685}{29.589} 
\emlineto{59.398}{29.878} 
\emmoveto{59.398}{29.868} 
\emlineto{59.110}{30.152} 
\emmoveto{59.110}{30.142} 
\emlineto{58.822}{30.434} 
\emmoveto{58.822}{30.424} 
\emlineto{58.533}{30.716} 
\emmoveto{58.533}{30.706} 
\emlineto{58.244}{30.998} 
\emmoveto{58.244}{30.988} 
\emlineto{57.955}{31.280} 
\emmoveto{57.955}{31.270} 
\emlineto{57.664}{31.562} 
\emmoveto{57.664}{31.552} 
\emlineto{57.374}{31.864} 
\emmoveto{57.374}{31.854} 
\emlineto{57.083}{32.146} 
\emmoveto{57.083}{32.136} 
\emlineto{56.792}{32.428} 
\emmoveto{56.792}{32.418} 
\emlineto{56.500}{32.730} 
\emmoveto{56.500}{32.720} 
\emlineto{56.208}{33.012} 
\emmoveto{56.208}{33.002} 
\emlineto{55.916}{33.314} 
\emmoveto{55.916}{33.304} 
\emlineto{55.623}{33.617} 
\emmoveto{55.623}{33.607} 
\emlineto{55.330}{33.899} 
\emmoveto{55.330}{33.889} 
\emlineto{55.037}{34.201} 
\emmoveto{55.037}{34.191} 
\emlineto{54.743}{34.483} 
\emmoveto{54.743}{34.473} 
\emlineto{54.449}{34.785} 
\emmoveto{54.449}{34.775} 
\emlineto{54.155}{35.087} 
\emmoveto{54.155}{35.077} 
\emlineto{53.861}{35.389} 
\emmoveto{53.861}{35.379} 
\emlineto{53.566}{35.671} 
\emmoveto{53.566}{35.661} 
\emlineto{53.271}{35.973} 
\emmoveto{53.271}{35.963} 
\emlineto{52.976}{36.275} 
\emmoveto{52.976}{36.265} 
\emlineto{52.680}{36.577} 
\emmoveto{52.680}{36.567} 
\emlineto{52.385}{36.880} 
\emmoveto{52.385}{36.870} 
\emlineto{52.089}{37.182} 
\emmoveto{52.089}{37.172} 
\emlineto{51.793}{37.484} 
\emmoveto{51.793}{37.474} 
\emlineto{51.497}{37.766} 
\emmoveto{51.497}{37.756} 
\emlineto{51.201}{38.068} 
\emmoveto{51.201}{38.058} 
\emlineto{50.904}{38.370} 
\emmoveto{50.904}{38.360} 
\emlineto{50.607}{38.672} 
\emmoveto{50.607}{38.662} 
\emlineto{50.310}{38.974} 
\emmoveto{50.310}{38.964} 
\emlineto{50.013}{39.277} 
\emmoveto{50.013}{39.267} 
\emlineto{49.716}{39.579} 
\emmoveto{49.716}{39.569} 
\emlineto{49.419}{39.881} 
\emmoveto{49.419}{39.871} 
\emlineto{49.121}{40.183} 
\emmoveto{49.121}{40.173} 
\emlineto{48.824}{40.505} 
\emmoveto{48.824}{40.495} 
\emlineto{48.526}{40.807} 
\emmoveto{48.526}{40.797} 
\emlineto{48.228}{41.109} 
\emmoveto{48.228}{41.099} 
\emlineto{47.930}{41.412} 
\emmoveto{47.930}{41.402} 
\emlineto{47.632}{41.714} 
\emmoveto{47.632}{41.704} 
\emlineto{47.333}{42.016} 
\emmoveto{47.333}{42.006} 
\emlineto{47.035}{42.318} 
\emmoveto{47.035}{42.308} 
\emlineto{46.736}{42.620} 
\emmoveto{46.736}{42.610} 
\emlineto{46.437}{42.922} 
\emmoveto{46.437}{42.912} 
\emlineto{46.139}{43.245} 
\emmoveto{46.139}{43.235} 
\emlineto{45.840}{43.547} 
\emmoveto{45.840}{43.537} 
\emlineto{45.541}{43.849} 
\emmoveto{45.541}{43.839} 
\emlineto{45.242}{44.151} 
\emmoveto{45.242}{44.141} 
\emlineto{44.942}{44.453} 
\emmoveto{44.942}{44.443} 
\emlineto{44.643}{44.775} 
\emmoveto{44.643}{44.765} 
\emlineto{44.344}{45.077} 
\emmoveto{44.344}{45.067} 
\emlineto{44.044}{45.380} 
\emmoveto{44.044}{45.370} 
\emlineto{43.744}{45.682} 
\emmoveto{43.744}{45.672} 
\emlineto{43.445}{46.004} 
\emmoveto{43.445}{45.994} 
\emlineto{43.145}{46.306} 
\emmoveto{43.145}{46.296} 
\emlineto{42.845}{46.608} 
\emmoveto{42.845}{46.598} 
\emlineto{42.545}{46.910} 
\emmoveto{42.545}{46.900} 
\emlineto{42.245}{47.233} 
\emmoveto{42.245}{47.223} 
\emlineto{41.945}{47.535} 
\emmoveto{41.945}{47.525} 
\emlineto{41.645}{47.837} 
\emmoveto{41.645}{47.827} 
\emlineto{41.344}{48.139} 
\emmoveto{41.344}{48.129} 
\emlineto{41.044}{48.461} 
\emmoveto{41.044}{48.451} 
\emlineto{40.743}{48.763} 
\emmoveto{40.743}{48.753} 
\emlineto{40.443}{49.066} 
\emmoveto{40.443}{49.056} 
\emlineto{40.142}{49.388} 
\emmoveto{40.142}{49.378} 
\emlineto{39.842}{49.690} 
\emmoveto{39.842}{49.680} 
\emlineto{39.541}{49.992} 
\emmoveto{39.541}{49.982} 
\emlineto{39.240}{50.314} 
\emmoveto{39.240}{50.304} 
\emlineto{38.939}{50.617} 
\emmoveto{38.939}{50.607} 
\emlineto{38.638}{50.919} 
\emmoveto{38.638}{50.909} 
\emlineto{38.337}{51.241} 
\emmoveto{38.337}{51.231} 
\emlineto{38.036}{51.543} 
\emmoveto{38.036}{51.533} 
\emlineto{37.735}{51.845} 
\emmoveto{37.735}{51.835} 
\emlineto{37.434}{52.168} 
\emmoveto{37.434}{52.158} 
\emlineto{37.133}{52.470} 
\emmoveto{37.133}{52.460} 
\emlineto{36.832}{52.772} 
\emmoveto{36.832}{52.762} 
\emlineto{36.530}{53.094} 
\emmoveto{36.530}{53.084} 
\emlineto{36.229}{53.396} 
\emmoveto{36.229}{53.386} 
\emlineto{35.927}{53.698} 
\emmoveto{35.927}{53.688} 
\emlineto{35.626}{54.021} 
\emmoveto{35.626}{54.011} 
\emlineto{35.324}{54.323} 
\emmoveto{35.324}{54.313} 
\emlineto{35.023}{54.625} 
\emmoveto{35.023}{54.615} 
\emlineto{34.721}{54.947} 
\emmoveto{34.721}{54.937} 
\emlineto{34.420}{55.249} 
\emmoveto{34.420}{55.239} 
\emlineto{34.118}{55.572} 
\emmoveto{34.118}{55.562} 
\emlineto{33.816}{55.874} 
\emmoveto{33.816}{55.864} 
\emlineto{33.514}{56.176} 
\emmoveto{33.514}{56.166} 
\emlineto{33.212}{56.498} 
\emmoveto{33.212}{56.488} 
\emlineto{32.910}{56.800} 
\emmoveto{32.910}{56.790} 
\emlineto{32.608}{57.122} 
\emmoveto{32.608}{57.112} 
\emlineto{32.306}{57.425} 
\emmoveto{32.306}{57.415} 
\emlineto{32.004}{57.727} 
\emmoveto{32.004}{57.717} 
\emlineto{31.702}{58.049} 
\emmoveto{31.702}{58.039} 
\emlineto{31.400}{58.351} 
\emmoveto{31.400}{58.341} 
\emlineto{31.098}{58.673} 
\emmoveto{31.098}{58.663} 
\emlineto{30.796}{58.976} 
\emmoveto{30.796}{58.966} 
\emlineto{30.493}{59.298} 
\emmoveto{30.493}{59.288} 
\emlineto{30.191}{59.600} 
\emmoveto{30.191}{59.590} 
\emlineto{29.889}{59.902} 
\emmoveto{29.889}{59.892} 
\emlineto{29.586}{60.224} 
\emmoveto{29.586}{60.214} 
\emlineto{29.284}{60.527} 
\emmoveto{29.284}{60.517} 
\emlineto{28.982}{60.849} 
\emmoveto{28.982}{60.839} 
\emlineto{28.679}{61.151} 
\emmoveto{28.679}{61.141} 
\emlineto{28.377}{61.473} 
\emmoveto{28.377}{61.463} 
\emlineto{28.074}{61.775} 
\emmoveto{28.074}{61.765} 
\emlineto{27.771}{62.098} 
\emmoveto{27.771}{62.088} 
\emlineto{27.469}{62.400} 
\emmoveto{27.469}{62.390} 
\emlineto{27.166}{62.702} 
\emmoveto{27.166}{62.692} 
\emlineto{26.863}{63.024} 
\emmoveto{26.863}{63.014} 
\emlineto{26.561}{63.326} 
\emmoveto{26.561}{63.316} 
\emlineto{26.258}{63.649} 
\emmoveto{26.258}{63.639} 
\emlineto{25.955}{63.951} 
\emmoveto{25.955}{63.941} 
\emlineto{25.652}{64.273} 
\emmoveto{25.652}{64.263} 
\emlineto{25.350}{64.575} 
\emmoveto{25.350}{64.565} 
\emlineto{25.047}{64.897} 
\emmoveto{25.047}{64.887} 
\emlineto{24.744}{65.199} 
\emmoveto{24.744}{65.189} 
\emlineto{24.441}{65.522} 
\emmoveto{24.441}{65.512} 
\emlineto{24.138}{65.824} 
\emmoveto{24.138}{65.814} 
\emlineto{23.835}{66.146} 
\emmoveto{23.835}{66.136} 
\emlineto{23.532}{66.448} 
\emmoveto{23.532}{66.438} 
\emlineto{23.229}{66.771} 
\emmoveto{23.229}{66.761} 
\emlineto{22.926}{67.073} 
\emmoveto{22.926}{67.063} 
\emlineto{22.623}{67.375} 
\emmoveto{22.623}{67.365} 
\emlineto{22.320}{67.697} 
\emmoveto{22.320}{67.687} 
\emlineto{22.016}{67.999} 
\emmoveto{22.016}{67.989} 
\emlineto{21.713}{68.322} 
\emmoveto{21.713}{68.312} 
\emlineto{21.410}{68.624} 
\emmoveto{21.410}{68.614} 
\emlineto{21.107}{68.946} 
\emmoveto{21.107}{68.936} 
\emlineto{20.804}{69.248} 
\emmoveto{20.804}{69.238} 
\emlineto{20.500}{69.570} 
\emmoveto{20.500}{69.560} 
\emlineto{20.197}{69.872} 
\emmoveto{20.197}{69.862} 
\emlineto{19.894}{70.195} 
\emmoveto{19.894}{70.185} 
\emlineto{19.590}{70.497} 
\emmoveto{19.590}{70.487} 
\emlineto{19.287}{70.819} 
\emmoveto{19.287}{70.809} 
\emlineto{18.984}{71.121} 
\emmoveto{18.984}{71.111} 
\emlineto{18.680}{71.444} 
\emmoveto{18.680}{71.434} 
\emlineto{18.377}{71.746} 
\emmoveto{18.377}{71.736} 
\emlineto{18.073}{72.068} 
\emmoveto{18.073}{72.058} 
\emlineto{17.770}{72.370} 
\emmoveto{17.770}{72.360} 
\emlineto{17.466}{72.692} 
\emmoveto{17.466}{72.682} 
\emlineto{17.163}{72.994} 
\emmoveto{17.163}{72.984} 
\emlineto{16.859}{73.317} 
\emmoveto{16.859}{73.307} 
\emlineto{16.556}{73.619} 
\emmoveto{16.556}{73.609} 
\emlineto{16.252}{73.941} 
\emmoveto{16.252}{73.931} 
\emlineto{15.949}{74.263} 
\emmoveto{15.949}{74.253} 
\emlineto{15.645}{74.566} 
\emmoveto{15.645}{74.556} 
\emlineto{15.341}{74.888} 
\emmoveto{15.341}{74.878} 
\emlineto{15.038}{75.190} 
\emmoveto{15.038}{75.180} 
\emlineto{14.734}{75.512} 
\emmoveto{14.734}{75.502} 
\emlineto{14.430}{75.814} 
\emmoveto{14.430}{75.804} 
\emlineto{14.127}{76.137} 
\emmoveto{14.127}{76.127} 
\emlineto{13.823}{76.439} 
\emmoveto{13.823}{76.429} 
\emlineto{13.519}{76.761} 
\emmoveto{13.519}{76.751} 
\emlineto{13.215}{77.063} 
\emmoveto{13.215}{77.053} 
\emlineto{12.911}{77.385} 
\emmoveto{12.911}{77.375} 
\emlineto{12.608}{77.688} 
\emmoveto{12.608}{77.678} 
\emlineto{12.304}{78.010} 
\emshow{24.980}{59.700}{2} 
\emmoveto{129.676}{22.085} 
\emlineto{129.356}{22.095} 
\emmoveto{129.356}{22.085} 
\emlineto{129.036}{22.095} 
\emmoveto{129.036}{22.085} 
\emlineto{128.716}{22.095} 
\emmoveto{128.716}{22.085} 
\emlineto{128.396}{22.095} 
\emmoveto{128.396}{22.085} 
\emlineto{128.076}{22.095} 
\emmoveto{128.076}{22.085} 
\emlineto{127.756}{22.095} 
\emmoveto{127.756}{22.085} 
\emlineto{127.436}{22.095} 
\emmoveto{127.436}{22.085} 
\emlineto{127.117}{22.095} 
\emmoveto{127.117}{22.085} 
\emlineto{126.797}{22.095} 
\emmoveto{126.797}{22.085} 
\emlineto{126.477}{22.095} 
\emmoveto{126.477}{22.085} 
\emlineto{126.157}{22.096} 
\emmoveto{126.157}{22.086} 
\emlineto{125.837}{22.096} 
\emmoveto{125.837}{22.086} 
\emlineto{125.517}{22.096} 
\emmoveto{125.517}{22.086} 
\emlineto{125.198}{22.096} 
\emmoveto{125.198}{22.086} 
\emlineto{124.878}{22.096} 
\emmoveto{124.878}{22.086} 
\emlineto{124.558}{22.096} 
\emmoveto{124.558}{22.086} 
\emlineto{124.238}{22.096} 
\emmoveto{124.238}{22.086} 
\emlineto{123.919}{22.096} 
\emmoveto{123.919}{22.086} 
\emlineto{123.599}{22.096} 
\emmoveto{123.599}{22.086} 
\emlineto{123.279}{22.096} 
\emmoveto{123.279}{22.086} 
\emlineto{122.960}{22.096} 
\emmoveto{122.960}{22.086} 
\emlineto{122.640}{22.096} 
\emmoveto{122.640}{22.086} 
\emlineto{122.321}{22.096} 
\emmoveto{122.321}{22.086} 
\emlineto{122.001}{22.096} 
\emmoveto{122.001}{22.086} 
\emlineto{121.681}{22.096} 
\emmoveto{121.681}{22.086} 
\emlineto{121.362}{22.097} 
\emmoveto{121.362}{22.087} 
\emlineto{121.042}{22.097} 
\emmoveto{121.042}{22.087} 
\emlineto{120.723}{22.097} 
\emmoveto{120.723}{22.087} 
\emlineto{120.403}{22.097} 
\emmoveto{120.403}{22.087} 
\emlineto{120.084}{22.097} 
\emmoveto{120.084}{22.087} 
\emlineto{119.765}{22.097} 
\emmoveto{119.765}{22.087} 
\emlineto{119.445}{22.097} 
\emmoveto{119.445}{22.087} 
\emlineto{119.126}{22.097} 
\emmoveto{119.126}{22.087} 
\emlineto{118.806}{22.098} 
\emmoveto{118.806}{22.088} 
\emlineto{118.487}{22.098} 
\emmoveto{118.487}{22.088} 
\emlineto{118.168}{22.098} 
\emmoveto{118.168}{22.088} 
\emlineto{117.848}{22.098} 
\emmoveto{117.848}{22.088} 
\emlineto{117.529}{22.098} 
\emmoveto{117.529}{22.088} 
\emlineto{117.210}{22.098} 
\emmoveto{117.210}{22.088} 
\emlineto{116.891}{22.098} 
\emmoveto{116.891}{22.088} 
\emlineto{116.572}{22.099} 
\emmoveto{116.572}{22.089} 
\emlineto{116.252}{22.099} 
\emmoveto{116.252}{22.089} 
\emlineto{115.933}{22.099} 
\emmoveto{115.933}{22.089} 
\emlineto{115.614}{22.099} 
\emmoveto{115.614}{22.089} 
\emlineto{115.295}{22.099} 
\emmoveto{115.295}{22.089} 
\emlineto{114.976}{22.100} 
\emmoveto{114.976}{22.090} 
\emlineto{114.657}{22.100} 
\emmoveto{114.657}{22.090} 
\emlineto{114.338}{22.100} 
\emmoveto{114.338}{22.090} 
\emlineto{114.019}{22.100} 
\emmoveto{114.019}{22.090} 
\emlineto{113.700}{22.101} 
\emmoveto{113.700}{22.091} 
\emlineto{113.381}{22.101} 
\emmoveto{113.381}{22.091} 
\emlineto{113.062}{22.101} 
\emmoveto{113.062}{22.091} 
\emlineto{112.743}{22.101} 
\emmoveto{112.743}{22.091} 
\emlineto{112.425}{22.102} 
\emmoveto{112.425}{22.092} 
\emlineto{112.106}{22.102} 
\emmoveto{112.106}{22.092} 
\emlineto{111.787}{22.102} 
\emmoveto{111.787}{22.092} 
\emlineto{111.468}{22.102} 
\emmoveto{111.468}{22.092} 
\emlineto{111.150}{22.103} 
\emmoveto{111.150}{22.093} 
\emlineto{110.831}{22.103} 
\emmoveto{110.831}{22.093} 
\emlineto{110.512}{22.103} 
\emmoveto{110.512}{22.093} 
\emlineto{110.194}{22.104} 
\emmoveto{110.194}{22.094} 
\emlineto{109.875}{22.104} 
\emmoveto{109.875}{22.094} 
\emlineto{109.556}{22.104} 
\emmoveto{109.556}{22.094} 
\emlineto{109.238}{22.105} 
\emmoveto{109.238}{22.095} 
\emlineto{108.919}{22.105} 
\emmoveto{108.919}{22.095} 
\emlineto{108.601}{22.105} 
\emmoveto{108.601}{22.095} 
\emlineto{108.283}{22.106} 
\emmoveto{108.283}{22.096} 
\emlineto{107.964}{22.106} 
\emmoveto{107.964}{22.096} 
\emlineto{107.646}{22.107} 
\emmoveto{107.646}{22.097} 
\emlineto{107.328}{22.107} 
\emmoveto{107.328}{22.097} 
\emlineto{107.009}{22.107} 
\emmoveto{107.009}{22.097} 
\emlineto{106.691}{22.108} 
\emmoveto{106.691}{22.098} 
\emlineto{106.373}{22.108} 
\emmoveto{106.373}{22.098} 
\emlineto{106.055}{22.109} 
\emmoveto{106.055}{22.099} 
\emlineto{105.737}{22.109} 
\emmoveto{105.737}{22.099} 
\emlineto{105.419}{22.110} 
\emmoveto{105.419}{22.100} 
\emlineto{105.101}{22.110} 
\emmoveto{105.101}{22.100} 
\emlineto{104.783}{22.111} 
\emmoveto{104.783}{22.101} 
\emlineto{104.465}{22.111} 
\emmoveto{104.465}{22.101} 
\emlineto{104.147}{22.112} 
\emmoveto{104.147}{22.102} 
\emlineto{103.829}{22.112} 
\emmoveto{103.829}{22.102} 
\emlineto{103.511}{22.113} 
\emmoveto{103.511}{22.103} 
\emlineto{103.193}{22.113} 
\emmoveto{103.193}{22.103} 
\emlineto{102.876}{22.114} 
\emmoveto{102.876}{22.104} 
\emlineto{102.558}{22.115} 
\emmoveto{102.558}{22.105} 
\emlineto{102.240}{22.115} 
\emmoveto{102.240}{22.105} 
\emlineto{101.923}{22.116} 
\emmoveto{101.923}{22.106} 
\emlineto{101.605}{22.117} 
\emmoveto{101.605}{22.107} 
\emlineto{101.288}{22.117} 
\emmoveto{101.288}{22.107} 
\emlineto{100.971}{22.118} 
\emmoveto{100.971}{22.108} 
\emlineto{100.653}{22.119} 
\emmoveto{100.653}{22.109} 
\emlineto{100.336}{22.119} 
\emmoveto{100.336}{22.109} 
\emlineto{100.019}{22.120} 
\emmoveto{100.019}{22.110} 
\emlineto{99.701}{22.121} 
\emmoveto{99.701}{22.111} 
\emlineto{99.384}{22.122} 
\emmoveto{99.384}{22.112} 
\emlineto{99.067}{22.123} 
\emmoveto{99.067}{22.113} 
\emlineto{98.750}{22.123} 
\emmoveto{98.750}{22.113} 
\emlineto{98.433}{22.124} 
\emmoveto{98.433}{22.114} 
\emlineto{98.116}{22.125} 
\emmoveto{98.116}{22.115} 
\emlineto{97.800}{22.126} 
\emmoveto{97.800}{22.116} 
\emlineto{97.483}{22.127} 
\emmoveto{97.483}{22.117} 
\emlineto{97.166}{22.128} 
\emmoveto{97.166}{22.118} 
\emlineto{96.849}{22.129} 
\emmoveto{96.849}{22.119} 
\emlineto{96.533}{22.130} 
\emmoveto{96.533}{22.120} 
\emlineto{96.216}{22.131} 
\emmoveto{96.216}{22.121} 
\emlineto{95.900}{22.132} 
\emmoveto{95.900}{22.122} 
\emlineto{95.583}{22.133} 
\emmoveto{95.583}{22.123} 
\emlineto{95.267}{22.134} 
\emmoveto{95.267}{22.124} 
\emlineto{94.951}{22.135} 
\emmoveto{94.951}{22.125} 
\emlineto{94.635}{22.137} 
\emmoveto{94.635}{22.127} 
\emlineto{94.319}{22.138} 
\emmoveto{94.319}{22.128} 
\emlineto{94.003}{22.139} 
\emmoveto{94.003}{22.129} 
\emlineto{93.687}{22.140} 
\emmoveto{93.687}{22.130} 
\emlineto{93.371}{22.142} 
\emmoveto{93.371}{22.132} 
\emlineto{93.055}{22.143} 
\emmoveto{93.055}{22.133} 
\emlineto{92.740}{22.144} 
\emmoveto{92.740}{22.134} 
\emlineto{92.424}{22.146} 
\emmoveto{92.424}{22.136} 
\emlineto{92.109}{22.147} 
\emmoveto{92.109}{22.137} 
\emlineto{91.793}{22.149} 
\emmoveto{91.793}{22.139} 
\emlineto{91.478}{22.150} 
\emmoveto{91.478}{22.140} 
\emlineto{91.163}{22.152} 
\emmoveto{91.163}{22.142} 
\emlineto{90.848}{22.154} 
\emmoveto{90.848}{22.144} 
\emlineto{90.533}{22.155} 
\emmoveto{90.533}{22.145} 
\emlineto{90.218}{22.157} 
\emmoveto{90.218}{22.147} 
\emlineto{89.903}{22.159} 
\emmoveto{89.903}{22.149} 
\emlineto{89.588}{22.161} 
\emmoveto{89.588}{22.151} 
\emlineto{89.274}{22.163} 
\emmoveto{89.274}{22.153} 
\emlineto{88.959}{22.165} 
\emmoveto{88.959}{22.155} 
\emlineto{88.645}{22.167} 
\emmoveto{88.645}{22.157} 
\emlineto{88.331}{22.169} 
\emmoveto{88.331}{22.159} 
\emlineto{88.017}{22.171} 
\emmoveto{88.017}{22.161} 
\emlineto{87.703}{22.173} 
\emmoveto{87.703}{22.163} 
\emlineto{87.389}{22.176} 
\emmoveto{87.389}{22.166} 
\emlineto{87.075}{22.178} 
\emmoveto{87.075}{22.168} 
\emlineto{86.762}{22.181} 
\emmoveto{86.762}{22.171} 
\emlineto{86.449}{22.183} 
\emmoveto{86.449}{22.173} 
\emlineto{86.135}{22.186} 
\emmoveto{86.135}{22.176} 
\emlineto{85.822}{22.189} 
\emmoveto{85.822}{22.179} 
\emlineto{85.509}{22.191} 
\emmoveto{85.509}{22.181} 
\emlineto{85.197}{22.194} 
\emmoveto{85.197}{22.184} 
\emlineto{84.884}{22.197} 
\emmoveto{84.884}{22.187} 
\emlineto{84.572}{22.201} 
\emmoveto{84.572}{22.191} 
\emlineto{84.259}{22.204} 
\emmoveto{84.259}{22.194} 
\emlineto{83.947}{22.207} 
\emmoveto{83.947}{22.197} 
\emlineto{83.635}{22.211} 
\emmoveto{83.635}{22.201} 
\emlineto{83.324}{22.215} 
\emmoveto{83.324}{22.205} 
\emlineto{83.012}{22.218} 
\emmoveto{83.012}{22.208} 
\emlineto{82.701}{22.222} 
\emmoveto{82.701}{22.212} 
\emlineto{82.390}{22.226} 
\emmoveto{82.390}{22.216} 
\emlineto{82.079}{22.231} 
\emmoveto{82.079}{22.221} 
\emlineto{81.769}{22.235} 
\emmoveto{81.769}{22.225} 
\emlineto{81.458}{22.240} 
\emmoveto{81.458}{22.230} 
\emlineto{81.148}{22.245} 
\emmoveto{81.148}{22.235} 
\emlineto{80.838}{22.250} 
\emmoveto{80.838}{22.240} 
\emlineto{80.529}{22.255} 
\emmoveto{80.529}{22.245} 
\emlineto{80.219}{22.261} 
\emmoveto{80.219}{22.251} 
\emlineto{79.910}{22.266} 
\emmoveto{79.910}{22.256} 
\emlineto{79.602}{22.272} 
\emmoveto{79.602}{22.262} 
\emlineto{79.293}{22.279} 
\emmoveto{79.293}{22.269} 
\emlineto{78.985}{22.285} 
\emmoveto{78.985}{22.275} 
\emlineto{78.678}{22.292} 
\emmoveto{78.678}{22.282} 
\emlineto{78.370}{22.300} 
\emmoveto{78.370}{22.290} 
\emlineto{78.063}{22.307} 
\emmoveto{78.063}{22.297} 
\emlineto{77.757}{22.315} 
\emmoveto{77.757}{22.305} 
\emlineto{77.451}{22.324} 
\emmoveto{77.451}{22.314} 
\emlineto{77.145}{22.333} 
\emmoveto{77.145}{22.323} 
\emlineto{76.839}{22.342} 
\emmoveto{76.839}{22.332} 
\emlineto{76.535}{22.352} 
\emmoveto{76.535}{22.342} 
\emlineto{76.230}{22.363} 
\emmoveto{76.230}{22.353} 
\emlineto{75.926}{22.374} 
\emmoveto{75.926}{22.364} 
\emlineto{75.623}{22.386} 
\emmoveto{75.623}{22.376} 
\emlineto{75.320}{22.398} 
\emmoveto{75.320}{22.388} 
\emlineto{75.018}{22.412} 
\emmoveto{75.018}{22.402} 
\emlineto{74.717}{22.426} 
\emmoveto{74.717}{22.416} 
\emlineto{74.416}{22.441} 
\emmoveto{74.416}{22.431} 
\emlineto{74.115}{22.457} 
\emmoveto{74.115}{22.447} 
\emlineto{73.816}{22.474} 
\emmoveto{73.816}{22.464} 
\emlineto{73.517}{22.492} 
\emmoveto{73.517}{22.482} 
\emlineto{73.219}{22.512} 
\emmoveto{73.219}{22.502} 
\emlineto{72.922}{22.533} 
\emmoveto{72.922}{22.523} 
\emlineto{72.626}{22.556} 
\emmoveto{72.626}{22.546} 
\emlineto{72.331}{22.581} 
\emmoveto{72.331}{22.571} 
\emlineto{72.036}{22.607} 
\emmoveto{72.036}{22.597} 
\emlineto{71.743}{22.636} 
\emmoveto{71.743}{22.626} 
\emlineto{71.451}{22.667} 
\emmoveto{71.451}{22.657} 
\emlineto{71.160}{22.701} 
\emmoveto{71.160}{22.691} 
\emlineto{70.870}{22.738} 
\emmoveto{70.870}{22.728} 
\emlineto{70.582}{22.778} 
\emmoveto{70.582}{22.768} 
\emlineto{70.294}{22.822} 
\emmoveto{70.294}{22.812} 
\emlineto{70.009}{22.869} 
\emmoveto{70.009}{22.859} 
\emlineto{69.724}{22.921} 
\emmoveto{69.724}{22.911} 
\emlineto{69.441}{22.979} 
\emmoveto{69.441}{22.969} 
\emlineto{69.159}{23.041} 
\emmoveto{69.159}{23.031} 
\emlineto{68.879}{23.110} 
\emmoveto{68.879}{23.100} 
\emlineto{68.601}{23.185} 
\emmoveto{68.601}{23.175} 
\emlineto{68.323}{23.268} 
\emmoveto{68.323}{23.258} 
\emlineto{68.048}{23.358} 
\emmoveto{68.048}{23.348} 
\emlineto{67.773}{23.457} 
\emmoveto{67.773}{23.447} 
\emlineto{67.500}{23.564} 
\emmoveto{67.500}{23.554} 
\emlineto{67.227}{23.681} 
\emmoveto{67.227}{23.671} 
\emlineto{66.956}{23.807} 
\emmoveto{66.956}{23.797} 
\emlineto{66.685}{23.944} 
\emmoveto{66.685}{23.934} 
\emlineto{66.415}{24.091} 
\emmoveto{66.415}{24.081} 
\emlineto{66.145}{24.247} 
\emmoveto{66.145}{24.237} 
\emlineto{65.875}{24.414} 
\emmoveto{65.875}{24.404} 
\emlineto{65.605}{24.591} 
\emmoveto{65.605}{24.581} 
\emlineto{65.335}{24.777} 
\emmoveto{65.335}{24.767} 
\emlineto{65.064}{24.973} 
\emmoveto{65.064}{24.963} 
\emlineto{64.793}{25.177} 
\emmoveto{64.793}{25.167} 
\emlineto{64.521}{25.389} 
\emmoveto{64.521}{25.379} 
\emlineto{64.248}{25.608} 
\emmoveto{64.248}{25.598} 
\emlineto{63.975}{25.835} 
\emmoveto{63.975}{25.825} 
\emlineto{63.701}{26.068} 
\emmoveto{63.701}{26.058} 
\emlineto{63.426}{26.307} 
\emmoveto{63.426}{26.297} 
\emlineto{63.150}{26.552} 
\emmoveto{63.150}{26.542} 
\emlineto{62.873}{26.802} 
\emmoveto{62.873}{26.792} 
\emlineto{62.596}{27.056} 
\emmoveto{62.596}{27.046} 
\emlineto{62.317}{27.315} 
\emmoveto{62.317}{27.305} 
\emlineto{62.038}{27.578} 
\emmoveto{62.038}{27.568} 
\emlineto{61.758}{27.844} 
\emmoveto{61.758}{27.834} 
\emlineto{61.477}{28.114} 
\emmoveto{61.477}{28.104} 
\emlineto{61.195}{28.387} 
\emmoveto{61.195}{28.377} 
\emlineto{60.913}{28.662} 
\emmoveto{60.913}{28.652} 
\emlineto{60.630}{28.940} 
\emmoveto{60.630}{28.930} 
\emlineto{60.346}{29.220} 
\emmoveto{60.346}{29.210} 
\emlineto{60.062}{29.503} 
\emmoveto{60.062}{29.493} 
\emlineto{59.777}{29.788} 
\emmoveto{59.777}{29.778} 
\emlineto{59.491}{30.074} 
\emmoveto{59.491}{30.064} 
\emlineto{59.205}{30.362} 
\emmoveto{59.205}{30.352} 
\emlineto{58.919}{30.652} 
\emmoveto{58.919}{30.642} 
\emlineto{58.631}{30.942} 
\emmoveto{58.631}{30.932} 
\emlineto{58.344}{31.236} 
\emmoveto{58.344}{31.226} 
\emlineto{58.055}{31.530} 
\emmoveto{58.055}{31.520} 
\emlineto{57.767}{31.824} 
\emmoveto{57.767}{31.814} 
\emlineto{57.478}{32.122} 
\emmoveto{57.478}{32.112} 
\emlineto{57.188}{32.418} 
\emmoveto{57.188}{32.408} 
\emlineto{56.898}{32.716} 
\emmoveto{56.898}{32.706} 
\emlineto{56.608}{33.016} 
\emmoveto{56.608}{33.006} 
\emlineto{56.318}{33.316} 
\emmoveto{56.318}{33.306} 
\emlineto{56.027}{33.619} 
\emmoveto{56.027}{33.609} 
\emlineto{55.735}{33.921} 
\emmoveto{55.735}{33.911} 
\emlineto{55.444}{34.223} 
\emmoveto{55.444}{34.213} 
\emlineto{55.152}{34.527} 
\emmoveto{55.152}{34.517} 
\emlineto{54.860}{34.831} 
\emmoveto{54.860}{34.821} 
\emlineto{54.567}{35.135} 
\emmoveto{54.567}{35.125} 
\emlineto{54.274}{35.441} 
\emmoveto{54.274}{35.431} 
\emlineto{53.981}{35.748} 
\emmoveto{53.981}{35.738} 
\emlineto{53.688}{36.054} 
\emmoveto{53.688}{36.044} 
\emlineto{53.394}{36.360} 
\emmoveto{53.394}{36.350} 
\emlineto{53.100}{36.668} 
\emmoveto{53.100}{36.658} 
\emlineto{52.806}{36.976} 
\emmoveto{52.806}{36.966} 
\emlineto{52.512}{37.284} 
\emmoveto{52.512}{37.274} 
\emlineto{52.218}{37.595} 
\emmoveto{52.218}{37.585} 
\emlineto{51.923}{37.903} 
\emmoveto{51.923}{37.893} 
\emlineto{51.628}{38.213} 
\emmoveto{51.628}{38.203} 
\emlineto{51.333}{38.523} 
\emmoveto{51.333}{38.513} 
\emlineto{51.038}{38.833} 
\emmoveto{51.038}{38.823} 
\emlineto{50.742}{39.144} 
\emmoveto{50.742}{39.134} 
\emlineto{50.447}{39.456} 
\emmoveto{50.447}{39.446} 
\emlineto{50.151}{39.768} 
\emmoveto{50.151}{39.758} 
\emlineto{49.855}{40.078} 
\emmoveto{49.855}{40.068} 
\emlineto{49.559}{40.390} 
\emmoveto{49.559}{40.380} 
\emlineto{49.263}{40.705} 
\emmoveto{49.263}{40.695} 
\emlineto{48.966}{41.017} 
\emmoveto{48.966}{41.007} 
\emlineto{48.670}{41.329} 
\emmoveto{48.670}{41.319} 
\emlineto{48.373}{41.643} 
\emmoveto{48.373}{41.633} 
\emlineto{48.076}{41.955} 
\emmoveto{48.076}{41.945} 
\emlineto{47.779}{42.270} 
\emmoveto{47.779}{42.260} 
\emlineto{47.482}{42.584} 
\emmoveto{47.482}{42.574} 
\emlineto{47.185}{42.898} 
\emmoveto{47.185}{42.888} 
\emlineto{46.887}{43.212} 
\emmoveto{46.887}{43.202} 
\emlineto{46.590}{43.527} 
\emmoveto{46.590}{43.517} 
\emlineto{46.292}{43.841} 
\emmoveto{46.292}{43.831} 
\emlineto{45.994}{44.157} 
\emmoveto{45.994}{44.147} 
\emlineto{45.697}{44.471} 
\emmoveto{45.697}{44.461} 
\emlineto{45.399}{44.787} 
\emmoveto{45.399}{44.777} 
\emlineto{45.101}{45.102} 
\emmoveto{45.101}{45.092} 
\emlineto{44.802}{45.418} 
\emmoveto{44.802}{45.408} 
\emlineto{44.504}{45.734} 
\emmoveto{44.504}{45.724} 
\emlineto{44.206}{46.050} 
\emmoveto{44.206}{46.040} 
\emlineto{43.907}{46.367} 
\emmoveto{43.907}{46.357} 
\emlineto{43.609}{46.683} 
\emmoveto{43.609}{46.673} 
\emlineto{43.310}{46.999} 
\emmoveto{43.310}{46.989} 
\emlineto{43.012}{47.315} 
\emmoveto{43.012}{47.305} 
\emlineto{42.713}{47.634} 
\emmoveto{42.713}{47.624} 
\emlineto{42.414}{47.950} 
\emmoveto{42.414}{47.940} 
\emlineto{42.115}{48.266} 
\emmoveto{42.115}{48.256} 
\emlineto{41.816}{48.584} 
\emmoveto{41.816}{48.574} 
\emlineto{41.517}{48.900} 
\emmoveto{41.517}{48.890} 
\emlineto{41.217}{49.219} 
\emmoveto{41.217}{49.209} 
\emlineto{40.918}{49.537} 
\emmoveto{40.918}{49.527} 
\emlineto{40.619}{49.853} 
\emmoveto{40.619}{49.843} 
\emlineto{40.319}{50.171} 
\emmoveto{40.319}{50.161} 
\emlineto{40.020}{50.490} 
\emmoveto{40.020}{50.480} 
\emlineto{39.720}{50.808} 
\emmoveto{39.720}{50.798} 
\emlineto{39.421}{51.126} 
\emmoveto{39.421}{51.116} 
\emlineto{39.121}{51.444} 
\emmoveto{39.121}{51.434} 
\emlineto{38.821}{51.763} 
\emmoveto{38.821}{51.753} 
\emlineto{38.521}{52.081} 
\emmoveto{38.521}{52.071} 
\emlineto{38.221}{52.399} 
\emmoveto{38.221}{52.389} 
\emlineto{37.921}{52.717} 
\emmoveto{37.921}{52.707} 
\emlineto{37.621}{53.036} 
\emmoveto{37.621}{53.026} 
\emlineto{37.321}{53.356} 
\emmoveto{37.321}{53.346} 
\emlineto{37.021}{53.674} 
\emmoveto{37.021}{53.664} 
\emlineto{36.721}{53.992} 
\emmoveto{36.721}{53.982} 
\emlineto{36.421}{54.313} 
\emmoveto{36.421}{54.303} 
\emlineto{36.120}{54.631} 
\emmoveto{36.120}{54.621} 
\emlineto{35.820}{54.951} 
\emmoveto{35.820}{54.941} 
\emlineto{35.520}{55.269} 
\emmoveto{35.520}{55.259} 
\emlineto{35.219}{55.590} 
\emmoveto{35.219}{55.580} 
\emlineto{34.919}{55.908} 
\emmoveto{34.919}{55.898} 
\emlineto{34.618}{56.228} 
\emmoveto{34.618}{56.218} 
\emlineto{34.317}{56.548} 
\emmoveto{34.317}{56.538} 
\emlineto{34.017}{56.869} 
\emmoveto{34.017}{56.859} 
\emlineto{33.716}{57.187} 
\emmoveto{33.716}{57.177} 
\emlineto{33.415}{57.507} 
\emmoveto{33.415}{57.497} 
\emlineto{33.114}{57.827} 
\emmoveto{33.114}{57.817} 
\emlineto{32.814}{58.148} 
\emmoveto{32.814}{58.138} 
\emlineto{32.513}{58.468} 
\emmoveto{32.513}{58.458} 
\emlineto{32.212}{58.786} 
\emmoveto{32.212}{58.776} 
\emlineto{31.911}{59.106} 
\emmoveto{31.911}{59.096} 
\emlineto{31.610}{59.427} 
\emmoveto{31.610}{59.417} 
\emlineto{31.309}{59.747} 
\emmoveto{31.309}{59.737} 
\emlineto{31.007}{60.067} 
\emmoveto{31.007}{60.057} 
\emlineto{30.706}{60.388} 
\emmoveto{30.706}{60.378} 
\emlineto{30.405}{60.710} 
\emmoveto{30.405}{60.700} 
\emlineto{30.104}{61.030} 
\emmoveto{30.104}{61.020} 
\emlineto{29.803}{61.350} 
\emmoveto{29.803}{61.340} 
\emlineto{29.501}{61.671} 
\emmoveto{29.501}{61.661} 
\emlineto{29.200}{61.991} 
\emmoveto{29.200}{61.981} 
\emlineto{28.899}{62.311} 
\emmoveto{28.899}{62.301} 
\emlineto{28.597}{62.633} 
\emmoveto{28.597}{62.623} 
\emlineto{28.296}{62.954} 
\emmoveto{28.296}{62.944} 
\emlineto{27.994}{63.274} 
\emmoveto{27.994}{63.264} 
\emlineto{27.693}{63.596} 
\emmoveto{27.693}{63.586} 
\emlineto{27.391}{63.916} 
\emmoveto{27.391}{63.906} 
\emlineto{27.089}{64.237} 
\emmoveto{27.089}{64.227} 
\emlineto{26.788}{64.559} 
\emmoveto{26.788}{64.549} 
\emlineto{26.486}{64.879} 
\emmoveto{26.486}{64.869} 
\emlineto{26.184}{65.199} 
\emmoveto{26.184}{65.189} 
\emlineto{25.883}{65.522} 
\emmoveto{25.883}{65.512} 
\emlineto{25.581}{65.842} 
\emmoveto{25.581}{65.832} 
\emlineto{25.279}{66.164} 
\emmoveto{25.279}{66.154} 
\emlineto{24.977}{66.485} 
\emmoveto{24.977}{66.475} 
\emlineto{24.675}{66.807} 
\emmoveto{24.675}{66.797} 
\emlineto{24.374}{67.129} 
\emmoveto{24.374}{67.119} 
\emlineto{24.072}{67.449} 
\emmoveto{24.072}{67.439} 
\emlineto{23.770}{67.772} 
\emmoveto{23.770}{67.762} 
\emlineto{23.468}{68.092} 
\emmoveto{23.468}{68.082} 
\emlineto{23.166}{68.414} 
\emmoveto{23.166}{68.404} 
\emlineto{22.864}{68.736} 
\emmoveto{22.864}{68.726} 
\emlineto{22.562}{69.057} 
\emmoveto{22.562}{69.047} 
\emlineto{22.260}{69.379} 
\emmoveto{22.260}{69.369} 
\emlineto{21.958}{69.701} 
\emmoveto{21.958}{69.691} 
\emlineto{21.655}{70.022} 
\emmoveto{21.655}{70.012} 
\emlineto{21.353}{70.344} 
\emmoveto{21.353}{70.334} 
\emlineto{21.051}{70.666} 
\emmoveto{21.051}{70.656} 
\emlineto{20.749}{70.988} 
\emmoveto{20.749}{70.978} 
\emlineto{20.447}{71.309} 
\emmoveto{20.447}{71.299} 
\emlineto{20.144}{71.631} 
\emmoveto{20.144}{71.621} 
\emlineto{19.842}{71.953} 
\emmoveto{19.842}{71.943} 
\emlineto{19.540}{72.275} 
\emmoveto{19.540}{72.265} 
\emlineto{19.238}{72.598} 
\emmoveto{19.238}{72.588} 
\emlineto{18.935}{72.920} 
\emmoveto{18.935}{72.910} 
\emlineto{18.633}{73.242} 
\emmoveto{18.633}{73.232} 
\emlineto{18.330}{73.562} 
\emmoveto{18.330}{73.552} 
\emlineto{18.028}{73.885} 
\emmoveto{18.028}{73.875} 
\emlineto{17.726}{74.207} 
\emmoveto{17.726}{74.197} 
\emlineto{17.423}{74.529} 
\emmoveto{17.423}{74.519} 
\emlineto{17.121}{74.852} 
\emmoveto{17.121}{74.842} 
\emlineto{16.818}{75.174} 
\emmoveto{16.818}{75.164} 
\emlineto{16.516}{75.496} 
\emmoveto{16.516}{75.486} 
\emlineto{16.213}{75.818} 
\emmoveto{16.213}{75.808} 
\emlineto{15.910}{76.141} 
\emmoveto{15.910}{76.131} 
\emlineto{15.608}{76.463} 
\emmoveto{15.608}{76.453} 
\emlineto{15.305}{76.785} 
\emmoveto{15.305}{76.775} 
\emlineto{15.003}{77.108} 
\emmoveto{15.003}{77.098} 
\emlineto{14.700}{77.430} 
\emmoveto{14.700}{77.420} 
\emlineto{14.397}{77.754} 
\emmoveto{14.397}{77.744} 
\emlineto{14.095}{78.076} 
\emmoveto{14.095}{78.066} 
\emlineto{13.792}{78.399} 
\emmoveto{13.792}{78.389} 
\emlineto{13.489}{78.721} 
\emmoveto{13.489}{78.711} 
\emlineto{13.186}{79.043} 
\emmoveto{13.186}{79.033} 
\emlineto{12.884}{79.365} 
\emmoveto{12.884}{79.355} 
\emlineto{12.581}{79.688} 
\emshow{24.980}{66.700}{1} 
\emshow{1.000}{10.000}{0.00e0} 
\emshow{1.000}{17.000}{3.48e1} 
\emshow{1.000}{24.000}{6.95e1} 
\emshow{1.000}{31.000}{1.04e2} 
\emshow{1.000}{38.000}{1.39e2} 
\emshow{1.000}{45.000}{1.74e2} 
\emshow{1.000}{52.000}{2.09e2} 
\emshow{1.000}{59.000}{2.43e2} 
\emshow{1.000}{66.000}{2.78e2} 
\emshow{1.000}{73.000}{3.13e2} 
\emshow{1.000}{80.000}{3.48e2} 
\emshow{12.000}{5.000}{-8.44e2} 
\emshow{23.800}{5.000}{-6.59e2} 
\emshow{35.600}{5.000}{-4.75e2} 
\emshow{47.400}{5.000}{-2.90e2} 
\emshow{59.200}{5.000}{-1.06e2} 
\emshow{71.000}{5.000}{7.82e1} 
\emshow{82.800}{5.000}{2.63e2} 
\emshow{94.600}{5.000}{4.47e2} 
\emshow{106.400}{5.000}{6.31e2} 
\emshow{118.200}{5.000}{8.16e2} 
\emshow{130.000}{5.000}{1.00e3}

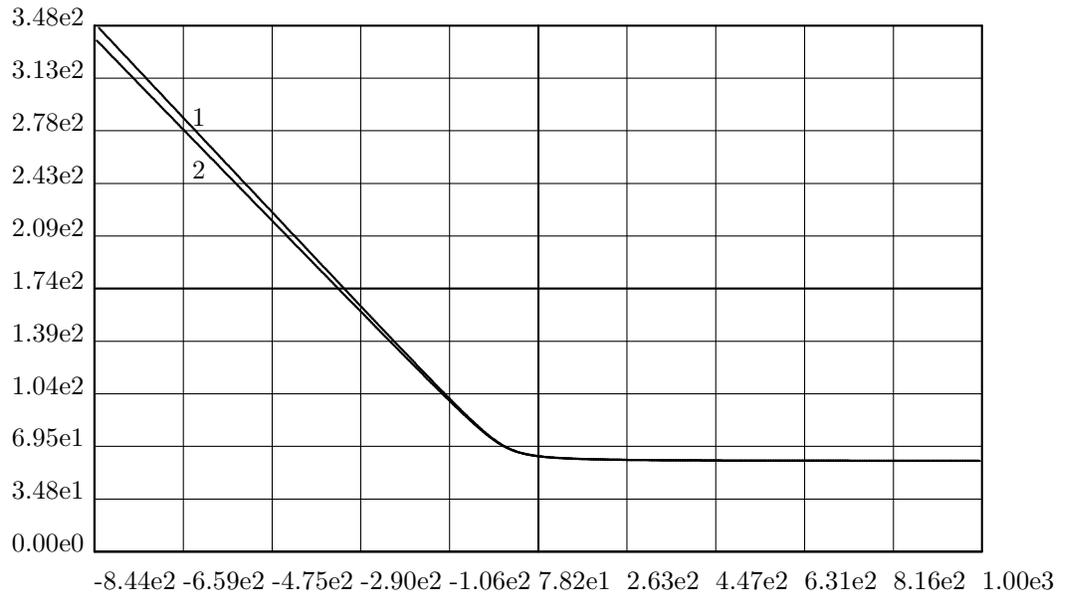
\centerline{ Fig. 1}

\newcount\numpoint                     
\newcount\numpointo                    
\numpoint=1 \numpointo=1               
\def\emmoveto#1#2{\offinterlineskip   
\hbox to 0 true cm{\vbox to 0          
true cm{\vskip - #2 true mm             
\hskip #1 true mm \special{em:point    
\the\numpoint}\vss}\hss}             
\numpointo=\numpoint                   
\global\advance \numpoint by 1}       
\def\emlineto#1#2{\offinterlineskip   
\hbox to 0 true cm{\vbox to 0          
true cm{\vskip - #2 true mm             
\hskip #1 true mm \special{em:point    
\the\numpoint}\vss}\hss}             
\special{em:line                        
\the\numpointo,\the\numpoint}        
\numpointo=\numpoint                   
\global\advance \numpoint by 1}       
\def\emshow#1#2#3{\offinterlineskip   
\hbox to 0 true cm{\vbox to 0          
true cm{\vskip - #2 true mm             
\hskip #1 true mm \vbox to 0           
true cm{\vss\hbox{#3\hss              
}}\vss}\hss}}                          
\special{em:linewidth 0.8pt}            

\vrule width 0 mm height                0 mm depth 90.000 true mm                

\special{em:linewidth 0.8pt} 
\emmoveto{130.000}{10.000} 
\emlineto{12.000}{10.000} 
\emlineto{12.000}{80.000} 
\emmoveto{71.000}{10.000} 
\emlineto{71.000}{80.000} 
\emmoveto{12.000}{45.000} 
\emlineto{130.000}{45.000} 
\emmoveto{130.000}{10.000} 
\emlineto{130.000}{80.000} 
\emlineto{12.000}{80.000} 
\emlineto{12.000}{10.000} 
\emlineto{130.000}{10.000} 
\special{em:linewidth 0.4pt} 
\emmoveto{12.000}{17.000} 
\emlineto{130.000}{17.000} 
\emmoveto{12.000}{24.000} 
\emlineto{130.000}{24.000} 
\emmoveto{12.000}{31.000} 
\emlineto{130.000}{31.000} 
\emmoveto{12.000}{38.000} 
\emlineto{130.000}{38.000} 
\emmoveto{12.000}{45.000} 
\emlineto{130.000}{45.000} 
\emmoveto{12.000}{52.000} 
\emlineto{130.000}{52.000} 
\emmoveto{12.000}{59.000} 
\emlineto{130.000}{59.000} 
\emmoveto{12.000}{66.000} 
\emlineto{130.000}{66.000} 
\emmoveto{12.000}{73.000} 
\emlineto{130.000}{73.000} 
\emmoveto{23.800}{10.000} 
\emlineto{23.800}{80.000} 
\emmoveto{35.600}{10.000} 
\emlineto{35.600}{80.000} 
\emmoveto{47.400}{10.000} 
\emlineto{47.400}{80.000} 
\emmoveto{59.200}{10.000} 
\emlineto{59.200}{80.000} 
\emmoveto{71.000}{10.000} 
\emlineto{71.000}{80.000} 
\emmoveto{82.800}{10.000} 
\emlineto{82.800}{80.000} 
\emmoveto{94.600}{10.000} 
\emlineto{94.600}{80.000} 
\emmoveto{106.400}{10.000} 
\emlineto{106.400}{80.000} 
\emmoveto{118.200}{10.000} 
\emlineto{118.200}{80.000} 
\special{em:linewidth 0.8pt} 
\emmoveto{15.010}{12.314} 
\emlineto{18.774}{15.217} 
\emmoveto{18.774}{15.207} 
\emlineto{24.047}{19.266} 
\emmoveto{24.047}{19.256} 
\emlineto{30.833}{24.473} 
\emmoveto{30.833}{24.463} 
\emlineto{39.136}{30.836} 
\emmoveto{39.136}{30.826} 
\emlineto{48.962}{38.357} 
\emmoveto{48.962}{38.347} 
\emlineto{60.316}{47.035} 
\emmoveto{60.316}{47.025} 
\emlineto{73.207}{56.870} 
\emmoveto{73.207}{56.860} 
\emlineto{87.642}{67.861} 
\emshow{72.180}{73.700}{R} 
\emmoveto{15.857}{12.314} 
\emlineto{20.681}{15.217} 
\emmoveto{20.681}{15.207} 
\emlineto{27.439}{19.266} 
\emmoveto{27.439}{19.256} 
\emlineto{36.135}{24.473} 
\emmoveto{36.135}{24.463} 
\emlineto{46.775}{30.836} 
\emmoveto{46.775}{30.826} 
\emlineto{59.366}{38.357} 
\emmoveto{59.366}{38.347} 
\emlineto{73.917}{47.035} 
\emmoveto{73.917}{47.025} 
\emlineto{90.436}{56.870} 
\emmoveto{90.436}{56.860} 
\emlineto{108.935}{67.861} 
\emshow{119.380}{73.700}{d=1} 
\emmoveto{15.421}{12.314} 
\emlineto{19.698}{15.217} 
\emmoveto{19.698}{15.207} 
\emlineto{25.691}{19.266} 
\emmoveto{25.691}{19.256} 
\emlineto{33.402}{24.473} 
\emmoveto{33.402}{24.463} 
\emlineto{42.838}{30.836} 
\emmoveto{42.838}{30.826} 
\emlineto{54.003}{38.357} 
\emmoveto{54.003}{38.347} 
\emlineto{66.906}{47.035} 
\emmoveto{66.906}{47.025} 
\emlineto{81.555}{56.870} 
\emmoveto{81.555}{56.860} 
\emlineto{97.959}{67.861} 
\emshow{107.580}{73.700}{d=2} 
\emmoveto{15.281}{12.314} 
\emlineto{19.384}{15.217} 
\emmoveto{19.384}{15.207} 
\emlineto{25.132}{19.266} 
\emmoveto{25.132}{19.256} 
\emlineto{32.528}{24.473} 
\emmoveto{32.528}{24.463} 
\emlineto{41.578}{30.836} 
\emmoveto{41.578}{30.826} 
\emlineto{52.287}{38.357} 
\emmoveto{52.287}{38.347} 
\emlineto{64.663}{47.035} 
\emmoveto{64.663}{47.025} 
\emlineto{78.713}{56.870} 
\emmoveto{78.713}{56.860} 
\emlineto{94.447}{67.861} 
\emshow{95.780}{73.700}{d=3} 
\emmoveto{15.212}{12.314} 
\emlineto{19.229}{15.217} 
\emmoveto{19.229}{15.207} 
\emlineto{24.856}{19.266} 
\emmoveto{24.856}{19.256} 
\emlineto{32.098}{24.473} 
\emmoveto{32.098}{24.463} 
\emlineto{40.958}{30.836} 
\emmoveto{40.958}{30.826} 
\emlineto{51.443}{38.357} 
\emmoveto{51.443}{38.347} 
\emlineto{63.559}{47.035} 
\emmoveto{63.559}{47.025} 
\emlineto{77.315}{56.870} 
\emmoveto{77.315}{56.860} 
\emlineto{92.719}{67.861} 
\emshow{83.980}{73.700}{d=4} 
\emshow{1.000}{10.000}{0.00e0} 
\emshow{1.000}{17.000}{1.21e1} 
\emshow{1.000}{24.000}{2.42e1} 
\emshow{1.000}{31.000}{3.63e1} 
\emshow{1.000}{38.000}{4.84e1} 
\emshow{1.000}{45.000}{6.05e1} 
\emshow{1.000}{52.000}{7.26e1} 
\emshow{1.000}{59.000}{8.47e1} 
\emshow{1.000}{66.000}{9.68e1} 
\emshow{1.000}{73.000}{1.09e2} 
\emshow{1.000}{80.000}{1.21e2} 
\emshow{12.000}{5.000}{0.00e0} 
\emshow{23.800}{5.000}{1.60e1} 
\emshow{35.600}{5.000}{3.20e1} 
\emshow{47.400}{5.000}{4.80e1} 
\emshow{59.200}{5.000}{6.40e1} 
\emshow{71.000}{5.000}{8.00e1} 
\emshow{82.800}{5.000}{9.60e1} 
\emshow{94.600}{5.000}{1.12e2} 
\emshow{106.400}{5.000}{1.28e2} 
\emshow{118.200}{5.000}{1.44e2} 
\emshow{130.000}{5.000}{1.60e2}

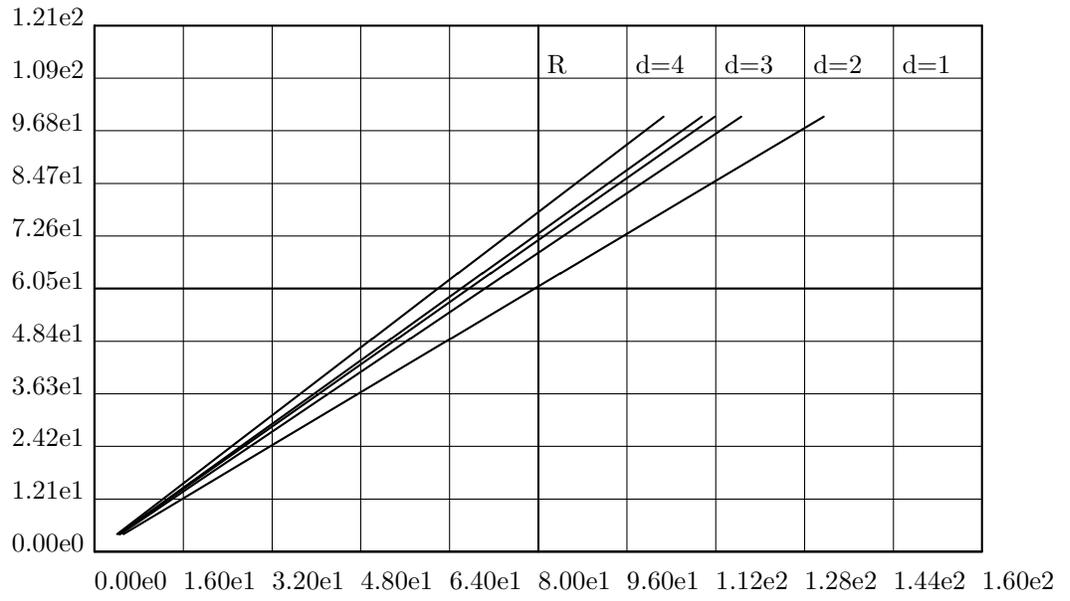
\centerline {\bf {Fig. 2}}
\newcount\numpoint                     
\newcount\numpointo                    
\numpoint=1 \numpointo=1               
\def\emmoveto#1#2{\offinterlineskip   
\hbox to 0 true cm{\vbox to 0          
true cm{\vskip - #2 true mm             
\hskip #1 true mm \special{em:point    
\the\numpoint}\vss}\hss}             
\numpointo=\numpoint                   
\global\advance \numpoint by 1}       
\def\emlineto#1#2{\offinterlineskip   
\hbox to 0 true cm{\vbox to 0          
true cm{\vskip - #2 true mm             
\hskip #1 true mm \special{em:point    
\the\numpoint}\vss}\hss}             
\special{em:line                        
\the\numpointo,\the\numpoint}        
\numpointo=\numpoint                   
\global\advance \numpoint by 1}       
\def\emshow#1#2#3{\offinterlineskip   
\hbox to 0 true cm{\vbox to 0          
true cm{\vskip - #2 true mm             
\hskip #1 true mm \vbox to 0           
true cm{\vss\hbox{#3\hss              
}}\vss}\hss}}                          
\special{em:linewidth 0.8pt}            

\vrule width 0 mm height                0 mm depth 90.000 true mm                

\special{em:linewidth 0.8pt} 
\emmoveto{130.000}{10.000} 
\emlineto{12.000}{10.000} 
\emlineto{12.000}{80.000} 
\emmoveto{71.000}{10.000} 
\emlineto{71.000}{80.000} 
\emmoveto{12.000}{45.000} 
\emlineto{130.000}{45.000} 
\emmoveto{130.000}{10.000} 
\emlineto{130.000}{80.000} 
\emlineto{12.000}{80.000} 
\emlineto{12.000}{10.000} 
\emlineto{130.000}{10.000} 
\special{em:linewidth 0.4pt} 
\emmoveto{12.000}{17.000} 
\emlineto{130.000}{17.000} 
\emmoveto{12.000}{24.000} 
\emlineto{130.000}{24.000} 
\emmoveto{12.000}{31.000} 
\emlineto{130.000}{31.000} 
\emmoveto{12.000}{38.000} 
\emlineto{130.000}{38.000} 
\emmoveto{12.000}{45.000} 
\emlineto{130.000}{45.000} 
\emmoveto{12.000}{52.000} 
\emlineto{130.000}{52.000} 
\emmoveto{12.000}{59.000} 
\emlineto{130.000}{59.000} 
\emmoveto{12.000}{66.000} 
\emlineto{130.000}{66.000} 
\emmoveto{12.000}{73.000} 
\emlineto{130.000}{73.000} 
\emmoveto{23.800}{10.000} 
\emlineto{23.800}{80.000} 
\emmoveto{35.600}{10.000} 
\emlineto{35.600}{80.000} 
\emmoveto{47.400}{10.000} 
\emlineto{47.400}{80.000} 
\emmoveto{59.200}{10.000} 
\emlineto{59.200}{80.000} 
\emmoveto{71.000}{10.000} 
\emlineto{71.000}{80.000} 
\emmoveto{82.800}{10.000} 
\emlineto{82.800}{80.000} 
\emmoveto{94.600}{10.000} 
\emlineto{94.600}{80.000} 
\emmoveto{106.400}{10.000} 
\emlineto{106.400}{80.000} 
\emmoveto{118.200}{10.000} 
\emlineto{118.200}{80.000} 
\special{em:linewidth 0.8pt} 
\emmoveto{129.648}{58.125} 
\emlineto{129.295}{58.135} 
\emmoveto{129.295}{58.125} 
\emlineto{128.943}{58.135} 
\emmoveto{128.943}{58.125} 
\emlineto{128.590}{58.135} 
\emmoveto{128.590}{58.125} 
\emlineto{128.237}{58.135} 
\emmoveto{128.237}{58.125} 
\emlineto{127.885}{58.135} 
\emmoveto{127.885}{58.125} 
\emlineto{127.532}{58.135} 
\emmoveto{127.532}{58.125} 
\emlineto{127.179}{58.135} 
\emmoveto{127.179}{58.125} 
\emlineto{126.826}{58.125} 
\emlineto{126.826}{58.135} 
\emmoveto{126.826}{58.125} 
\emlineto{126.473}{58.124} 
\emlineto{126.473}{58.134} 
\emmoveto{126.473}{58.124} 
\emlineto{126.120}{58.134} 
\emmoveto{126.120}{58.124} 
\emlineto{125.766}{58.134} 
\emmoveto{125.766}{58.124} 
\emlineto{125.413}{58.124} 
\emlineto{125.413}{58.134} 
\emmoveto{125.413}{58.124} 
\emlineto{125.059}{58.124} 
\emlineto{125.059}{58.134} 
\emmoveto{125.059}{58.124} 
\emlineto{124.706}{58.134} 
\emmoveto{124.706}{58.124} 
\emlineto{124.352}{58.124} 
\emlineto{124.352}{58.134} 
\emmoveto{124.352}{58.124} 
\emlineto{123.998}{58.123} 
\emlineto{123.998}{58.133} 
\emmoveto{123.998}{58.123} 
\emlineto{123.645}{58.123} 
\emlineto{123.645}{58.133} 
\emmoveto{123.645}{58.123} 
\emlineto{123.291}{58.123} 
\emlineto{123.291}{58.133} 
\emmoveto{123.291}{58.123} 
\emlineto{122.937}{58.123} 
\emlineto{122.937}{58.133} 
\emmoveto{122.937}{58.123} 
\emlineto{122.583}{58.122} 
\emlineto{122.583}{58.132} 
\emmoveto{122.583}{58.122} 
\emlineto{122.228}{58.122} 
\emlineto{122.228}{58.132} 
\emmoveto{122.228}{58.122} 
\emlineto{121.874}{58.122} 
\emlineto{121.874}{58.132} 
\emmoveto{121.874}{58.122} 
\emlineto{121.520}{58.122} 
\emlineto{121.520}{58.132} 
\emmoveto{121.520}{58.122} 
\emlineto{121.165}{58.122} 
\emlineto{121.165}{58.132} 
\emmoveto{121.165}{58.122} 
\emlineto{120.811}{58.122} 
\emlineto{120.811}{58.132} 
\emmoveto{120.811}{58.122} 
\emlineto{120.456}{58.131} 
\emmoveto{120.456}{58.121} 
\emlineto{120.101}{58.121} 
\emlineto{120.101}{58.131} 
\emmoveto{120.101}{58.121} 
\emlineto{119.746}{58.121} 
\emlineto{119.746}{58.131} 
\emmoveto{119.746}{58.121} 
\emlineto{119.391}{58.130} 
\emmoveto{119.391}{58.120} 
\emlineto{119.036}{58.130} 
\emmoveto{119.036}{58.120} 
\emlineto{118.681}{58.119} 
\emlineto{118.681}{58.129} 
\emmoveto{118.681}{58.119} 
\emlineto{118.326}{58.119} 
\emlineto{118.326}{58.129} 
\emmoveto{118.326}{58.119} 
\emlineto{117.970}{58.129} 
\emmoveto{117.970}{58.119} 
\emlineto{117.615}{58.128} 
\emmoveto{117.615}{58.118} 
\emlineto{117.259}{58.128} 
\emmoveto{117.259}{58.118} 
\emlineto{116.903}{58.128} 
\emmoveto{116.903}{58.118} 
\emlineto{116.548}{58.127} 
\emmoveto{116.548}{58.117} 
\emlineto{116.192}{58.127} 
\emmoveto{116.192}{58.117} 
\emlineto{115.836}{58.126} 
\emmoveto{115.836}{58.116} 
\emlineto{115.479}{58.126} 
\emmoveto{115.479}{58.116} 
\emlineto{115.123}{58.125} 
\emmoveto{115.123}{58.115} 
\emlineto{114.767}{58.125} 
\emmoveto{114.767}{58.115} 
\emlineto{114.410}{58.125} 
\emmoveto{114.410}{58.115} 
\emlineto{114.054}{58.124} 
\emmoveto{114.054}{58.114} 
\emlineto{113.697}{58.123} 
\emmoveto{113.697}{58.113} 
\emlineto{113.340}{58.123} 
\emmoveto{113.340}{58.113} 
\emlineto{112.983}{58.122} 
\emmoveto{112.983}{58.112} 
\emlineto{112.626}{58.122} 
\emmoveto{112.626}{58.112} 
\emlineto{112.269}{58.121} 
\emmoveto{112.269}{58.111} 
\emlineto{111.912}{58.121} 
\emmoveto{111.912}{58.111} 
\emlineto{111.555}{58.120} 
\emmoveto{111.555}{58.110} 
\emlineto{111.197}{58.119} 
\emmoveto{111.197}{58.109} 
\emlineto{110.840}{58.118} 
\emmoveto{110.840}{58.108} 
\emlineto{110.482}{58.118} 
\emmoveto{110.482}{58.108} 
\emlineto{110.124}{58.117} 
\emmoveto{110.124}{58.107} 
\emlineto{109.766}{58.117} 
\emmoveto{109.766}{58.107} 
\emlineto{109.408}{58.116} 
\emmoveto{109.408}{58.106} 
\emlineto{109.050}{58.115} 
\emmoveto{109.050}{58.105} 
\emlineto{108.691}{58.114} 
\emmoveto{108.691}{58.104} 
\emlineto{108.333}{58.114} 
\emmoveto{108.333}{58.104} 
\emlineto{107.974}{58.113} 
\emmoveto{107.974}{58.103} 
\emlineto{107.616}{58.112} 
\emmoveto{107.616}{58.102} 
\emlineto{107.257}{58.111} 
\emmoveto{107.257}{58.101} 
\emlineto{106.898}{58.110} 
\emmoveto{106.898}{58.101} 
\emlineto{106.539}{58.110} 
\emmoveto{106.539}{58.100} 
\emlineto{106.179}{58.109} 
\emmoveto{106.179}{58.099} 
\emlineto{105.820}{58.108} 
\emmoveto{105.820}{58.098} 
\emlineto{105.461}{58.107} 
\emmoveto{105.461}{58.097} 
\emlineto{105.101}{58.106} 
\emmoveto{105.101}{58.096} 
\emlineto{104.741}{58.105} 
\emmoveto{104.741}{58.095} 
\emlineto{104.381}{58.104} 
\emmoveto{104.381}{58.094} 
\emlineto{104.021}{58.103} 
\emmoveto{104.021}{58.093} 
\emlineto{103.661}{58.102} 
\emmoveto{103.661}{58.092} 
\emlineto{103.301}{58.101} 
\emmoveto{103.301}{58.091} 
\emlineto{102.940}{58.100} 
\emmoveto{102.940}{58.090} 
\emlineto{102.580}{58.099} 
\emmoveto{102.580}{58.089} 
\emlineto{102.219}{58.098} 
\emmoveto{102.219}{58.088} 
\emlineto{101.858}{58.097} 
\emmoveto{101.858}{58.087} 
\emlineto{101.497}{58.096} 
\emmoveto{101.497}{58.086} 
\emlineto{101.136}{58.094} 
\emmoveto{101.136}{58.084} 
\emlineto{100.775}{58.093} 
\emmoveto{100.775}{58.083} 
\emlineto{100.413}{58.092} 
\emmoveto{100.413}{58.082} 
\emlineto{100.052}{58.091} 
\emmoveto{100.052}{58.081} 
\emlineto{99.690}{58.090} 
\emmoveto{99.690}{58.080} 
\emlineto{99.328}{58.088} 
\emmoveto{99.328}{58.078} 
\emlineto{98.966}{58.087} 
\emmoveto{98.966}{58.077} 
\emlineto{98.604}{58.086} 
\emmoveto{98.604}{58.076} 
\emlineto{98.241}{58.084} 
\emmoveto{98.241}{58.074} 
\emlineto{97.879}{58.083} 
\emmoveto{97.879}{58.073} 
\emlineto{97.516}{58.082} 
\emmoveto{97.516}{58.072} 
\emlineto{97.153}{58.080} 
\emmoveto{97.153}{58.070} 
\emlineto{96.790}{58.079} 
\emmoveto{96.790}{58.069} 
\emlineto{96.427}{58.077} 
\emmoveto{96.427}{58.067} 
\emlineto{96.064}{58.076} 
\emmoveto{96.064}{58.066} 
\emlineto{95.700}{58.074} 
\emmoveto{95.700}{58.064} 
\emlineto{95.337}{58.073} 
\emmoveto{95.337}{58.063} 
\emlineto{94.973}{58.071} 
\emmoveto{94.973}{58.061} 
\emlineto{94.609}{58.070} 
\emmoveto{94.609}{58.060} 
\emlineto{94.245}{58.068} 
\emmoveto{94.245}{58.058} 
\emlineto{93.880}{58.066} 
\emmoveto{93.880}{58.056} 
\emlineto{93.516}{58.065} 
\emmoveto{93.516}{58.055} 
\emlineto{93.151}{58.063} 
\emmoveto{93.151}{58.053} 
\emlineto{92.786}{58.062} 
\emmoveto{92.786}{58.052} 
\emlineto{92.421}{58.060} 
\emmoveto{92.421}{58.050} 
\emlineto{92.056}{58.058} 
\emmoveto{92.056}{58.048} 
\emlineto{91.691}{58.056} 
\emmoveto{91.691}{58.046} 
\emlineto{91.325}{58.054} 
\emmoveto{91.325}{58.044} 
\emlineto{90.959}{58.052} 
\emmoveto{90.959}{58.042} 
\emlineto{90.593}{58.051} 
\emmoveto{90.593}{58.041} 
\emlineto{90.227}{58.048} 
\emmoveto{90.227}{58.038} 
\emlineto{89.861}{58.047} 
\emmoveto{89.861}{58.037} 
\emlineto{89.494}{58.044} 
\emmoveto{89.494}{58.034} 
\emlineto{89.128}{58.043} 
\emmoveto{89.128}{58.033} 
\emlineto{88.761}{58.040} 
\emmoveto{88.761}{58.031} 
\emlineto{88.394}{58.038} 
\emmoveto{88.394}{58.028} 
\emlineto{88.026}{58.036} 
\emmoveto{88.026}{58.026} 
\emlineto{87.659}{58.034} 
\emmoveto{87.659}{58.024} 
\emlineto{87.291}{58.032} 
\emmoveto{87.291}{58.022} 
\emlineto{86.923}{58.030} 
\emmoveto{86.923}{58.020} 
\emlineto{86.555}{58.027} 
\emmoveto{86.555}{58.017} 
\emlineto{86.187}{58.025} 
\emmoveto{86.187}{58.015} 
\emlineto{85.818}{58.023} 
\emmoveto{85.818}{58.013} 
\emlineto{85.449}{58.020} 
\emmoveto{85.449}{58.010} 
\emlineto{85.080}{58.018} 
\emmoveto{85.080}{58.008} 
\emlineto{84.711}{58.016} 
\emmoveto{84.711}{58.006} 
\emlineto{84.342}{58.013} 
\emmoveto{84.342}{58.003} 
\emlineto{83.972}{58.010} 
\emmoveto{83.972}{58.000} 
\emlineto{83.602}{58.008} 
\emmoveto{83.602}{57.998} 
\emlineto{83.232}{58.005} 
\emmoveto{83.232}{57.995} 
\emlineto{82.862}{58.002} 
\emmoveto{82.862}{57.992} 
\emlineto{82.491}{58.000} 
\emmoveto{82.491}{57.990} 
\emlineto{82.121}{57.997} 
\emmoveto{82.121}{57.987} 
\emlineto{81.750}{57.994} 
\emmoveto{81.750}{57.984} 
\emlineto{81.378}{57.991} 
\emmoveto{81.378}{57.982} 
\emlineto{81.007}{57.988} 
\emmoveto{81.007}{57.978} 
\emlineto{80.635}{57.985} 
\emmoveto{80.635}{57.975} 
\emlineto{80.263}{57.982} 
\emmoveto{80.263}{57.972} 
\emlineto{79.891}{57.980} 
\emmoveto{79.891}{57.970} 
\emlineto{79.518}{57.977} 
\emmoveto{79.518}{57.967} 
\emlineto{79.146}{57.973} 
\emmoveto{79.146}{57.963} 
\emlineto{78.772}{57.970} 
\emmoveto{78.772}{57.960} 
\emlineto{78.399}{57.967} 
\emmoveto{78.399}{57.957} 
\emlineto{78.026}{57.964} 
\emmoveto{78.026}{57.954} 
\emlineto{77.652}{57.960} 
\emmoveto{77.652}{57.950} 
\emlineto{77.278}{57.957} 
\emmoveto{77.278}{57.947} 
\emlineto{76.903}{57.953} 
\emmoveto{76.903}{57.943} 
\emlineto{76.529}{57.950} 
\emmoveto{76.529}{57.940} 
\emlineto{76.154}{57.946} 
\emmoveto{76.154}{57.936} 
\emlineto{75.779}{57.943} 
\emmoveto{75.779}{57.933} 
\emlineto{75.403}{57.939} 
\emmoveto{75.403}{57.929} 
\emlineto{75.027}{57.936} 
\emmoveto{75.027}{57.926} 
\emlineto{74.651}{57.932} 
\emmoveto{74.651}{57.922} 
\emlineto{74.275}{57.928} 
\emmoveto{74.275}{57.918} 
\emlineto{73.898}{57.924} 
\emmoveto{73.898}{57.914} 
\emlineto{73.521}{57.920} 
\emmoveto{73.521}{57.910} 
\emlineto{73.144}{57.916} 
\emmoveto{73.144}{57.906} 
\emlineto{72.766}{57.911} 
\emmoveto{72.766}{57.901} 
\emlineto{72.389}{57.908} 
\emmoveto{72.389}{57.898} 
\emlineto{72.010}{57.903} 
\emmoveto{72.010}{57.893} 
\emlineto{71.632}{57.899} 
\emmoveto{71.632}{57.889} 
\emlineto{71.253}{57.894} 
\emmoveto{71.253}{57.884} 
\emlineto{70.874}{57.890} 
\emmoveto{70.874}{57.880} 
\emlineto{70.494}{57.886} 
\emmoveto{70.494}{57.876} 
\emlineto{70.114}{57.881} 
\emmoveto{70.114}{57.871} 
\emlineto{69.734}{57.876} 
\emmoveto{69.734}{57.866} 
\emlineto{69.353}{57.871} 
\emmoveto{69.353}{57.861} 
\emlineto{68.973}{57.866} 
\emmoveto{68.973}{57.856} 
\emlineto{68.591}{57.862} 
\emmoveto{68.591}{57.852} 
\emlineto{68.210}{57.857} 
\emmoveto{68.210}{57.847} 
\emlineto{67.828}{57.852} 
\emmoveto{67.828}{57.842} 
\emlineto{67.445}{57.846} 
\emmoveto{67.445}{57.836} 
\emlineto{67.062}{57.841} 
\emmoveto{67.062}{57.831} 
\emlineto{66.679}{57.836} 
\emmoveto{66.679}{57.826} 
\emlineto{66.296}{57.830} 
\emmoveto{66.296}{57.820} 
\emlineto{65.912}{57.824} 
\emmoveto{65.912}{57.814} 
\emlineto{65.527}{57.819} 
\emmoveto{65.527}{57.809} 
\emlineto{65.143}{57.813} 
\emmoveto{65.143}{57.803} 
\emlineto{64.757}{57.807} 
\emmoveto{64.757}{57.797} 
\emlineto{64.372}{57.801} 
\emmoveto{64.372}{57.791} 
\emlineto{63.986}{57.795} 
\emmoveto{63.986}{57.785} 
\emlineto{63.599}{57.789} 
\emmoveto{63.599}{57.779} 
\emlineto{63.212}{57.783} 
\emmoveto{63.212}{57.773} 
\emlineto{62.825}{57.776} 
\emmoveto{62.825}{57.766} 
\emlineto{62.437}{57.770} 
\emmoveto{62.437}{57.760} 
\emlineto{62.049}{57.763} 
\emmoveto{62.049}{57.753} 
\emlineto{61.660}{57.757} 
\emmoveto{61.660}{57.747} 
\emlineto{61.271}{57.750} 
\emmoveto{61.271}{57.740} 
\emlineto{60.881}{57.743} 
\emmoveto{60.881}{57.733} 
\emlineto{60.491}{57.736} 
\emmoveto{60.491}{57.726} 
\emlineto{60.100}{57.728} 
\emmoveto{60.100}{57.718} 
\emlineto{59.709}{57.721} 
\emmoveto{59.709}{57.711} 
\emlineto{59.318}{57.713} 
\emmoveto{59.318}{57.703} 
\emlineto{58.925}{57.706} 
\emmoveto{58.925}{57.696} 
\emlineto{58.533}{57.698} 
\emmoveto{58.533}{57.688} 
\emlineto{58.139}{57.690} 
\emmoveto{58.139}{57.680} 
\emlineto{57.746}{57.682} 
\emmoveto{57.746}{57.672} 
\emlineto{57.351}{57.673} 
\emmoveto{57.351}{57.663} 
\emlineto{56.956}{57.665} 
\emmoveto{56.956}{57.655} 
\emlineto{56.561}{57.656} 
\emmoveto{56.561}{57.646} 
\emlineto{56.165}{57.648} 
\emmoveto{56.165}{57.638} 
\emlineto{55.768}{57.639} 
\emmoveto{55.768}{57.629} 
\emlineto{55.370}{57.630} 
\emmoveto{55.370}{57.620} 
\emlineto{54.972}{57.621} 
\emmoveto{54.972}{57.611} 
\emlineto{54.574}{57.611} 
\emmoveto{54.574}{57.601} 
\emlineto{54.175}{57.601} 
\emmoveto{54.175}{57.591} 
\emlineto{53.775}{57.591} 
\emmoveto{53.775}{57.581} 
\emlineto{53.374}{57.581} 
\emmoveto{53.374}{57.571} 
\emlineto{52.973}{57.571} 
\emmoveto{52.973}{57.561} 
\emlineto{52.571}{57.560} 
\emmoveto{52.571}{57.550} 
\emlineto{52.168}{57.550} 
\emmoveto{52.168}{57.540} 
\emlineto{51.765}{57.539} 
\emmoveto{51.765}{57.529} 
\emlineto{51.361}{57.527} 
\emmoveto{51.361}{57.517} 
\emlineto{50.956}{57.516} 
\emmoveto{50.956}{57.506} 
\emlineto{50.550}{57.504} 
\emmoveto{50.550}{57.494} 
\emlineto{50.144}{57.492} 
\emmoveto{50.144}{57.482} 
\emlineto{49.736}{57.480} 
\emmoveto{49.736}{57.470} 
\emlineto{49.328}{57.467} 
\emmoveto{49.328}{57.457} 
\emlineto{48.919}{57.455} 
\emmoveto{48.919}{57.445} 
\emlineto{48.510}{57.442} 
\emmoveto{48.510}{57.432} 
\emlineto{48.099}{57.428} 
\emmoveto{48.099}{57.418} 
\emlineto{47.687}{57.414} 
\emmoveto{47.687}{57.404} 
\emlineto{47.275}{57.400} 
\emmoveto{47.275}{57.390} 
\emlineto{46.862}{57.386} 
\emmoveto{46.862}{57.376} 
\emlineto{46.447}{57.372} 
\emmoveto{46.447}{57.362} 
\emlineto{46.032}{57.356} 
\emmoveto{46.032}{57.346} 
\emlineto{45.615}{57.341} 
\emmoveto{45.615}{57.331} 
\emlineto{45.198}{57.325} 
\emmoveto{45.198}{57.315} 
\emlineto{44.779}{57.309} 
\emmoveto{44.779}{57.299} 
\emlineto{44.360}{57.292} 
\emmoveto{44.360}{57.282} 
\emlineto{43.939}{57.275} 
\emmoveto{43.939}{57.265} 
\emlineto{43.517}{57.257} 
\emmoveto{43.517}{57.247} 
\emlineto{43.094}{57.239} 
\emmoveto{43.094}{57.229} 
\emlineto{42.670}{57.221} 
\emmoveto{42.670}{57.211} 
\emlineto{42.244}{57.202} 
\emmoveto{42.244}{57.192} 
\emlineto{41.818}{57.182} 
\emmoveto{41.818}{57.172} 
\emlineto{41.390}{57.162} 
\emmoveto{41.390}{57.152} 
\emlineto{40.960}{57.141} 
\emmoveto{40.960}{57.131} 
\emlineto{40.529}{57.120} 
\emmoveto{40.529}{57.110} 
\emlineto{40.097}{57.099} 
\emmoveto{40.097}{57.089} 
\emlineto{39.663}{57.076} 
\emmoveto{39.663}{57.066} 
\emlineto{39.228}{57.053} 
\emmoveto{39.228}{57.043} 
\emlineto{38.791}{57.029} 
\emmoveto{38.791}{57.019} 
\emlineto{38.352}{57.004} 
\emmoveto{38.352}{56.994} 
\emlineto{37.912}{56.979} 
\emmoveto{37.912}{56.969} 
\emlineto{37.470}{56.952} 
\emmoveto{37.470}{56.942} 
\emlineto{37.026}{56.925} 
\emmoveto{37.026}{56.915} 
\emlineto{36.581}{56.897} 
\emmoveto{36.581}{56.887} 
\emlineto{36.133}{56.868} 
\emmoveto{36.133}{56.858} 
\emlineto{35.683}{56.838} 
\emmoveto{35.683}{56.828} 
\emlineto{35.231}{56.807} 
\emmoveto{35.231}{56.797} 
\emlineto{34.778}{56.774} 
\emmoveto{34.778}{56.764} 
\emlineto{34.321}{56.741} 
\emmoveto{34.321}{56.731} 
\emlineto{33.863}{56.706} 
\emmoveto{33.863}{56.696} 
\emlineto{33.402}{56.670} 
\emmoveto{33.402}{56.660} 
\emlineto{32.938}{56.633} 
\emmoveto{32.938}{56.623} 
\emlineto{32.472}{56.594} 
\emmoveto{32.472}{56.584} 
\emlineto{32.003}{56.553} 
\emmoveto{32.003}{56.543} 
\emlineto{31.531}{56.511} 
\emmoveto{31.531}{56.501} 
\emlineto{31.056}{56.466} 
\emmoveto{31.056}{56.456} 
\emlineto{30.577}{56.420} 
\emmoveto{30.577}{56.410} 
\emlineto{30.096}{56.371} 
\emmoveto{30.096}{56.361} 
\emlineto{29.610}{56.320} 
\emmoveto{29.610}{56.310} 
\emlineto{29.121}{56.267} 
\emmoveto{29.121}{56.257} 
\emlineto{28.628}{56.211} 
\emmoveto{28.628}{56.201} 
\emlineto{28.131}{56.152} 
\emmoveto{28.131}{56.142} 
\emlineto{27.629}{56.090} 
\emmoveto{27.629}{56.080} 
\emlineto{27.122}{56.024} 
\emmoveto{27.122}{56.014} 
\emlineto{26.611}{55.955} 
\emmoveto{26.611}{55.945} 
\emlineto{26.093}{55.881} 
\emmoveto{26.093}{55.871} 
\emlineto{25.570}{55.802} 
\emmoveto{25.570}{55.792} 
\emlineto{25.041}{55.718} 
\emmoveto{25.041}{55.708} 
\emlineto{24.504}{55.628} 
\emmoveto{24.504}{55.618} 
\emlineto{23.960}{55.532} 
\emmoveto{23.960}{55.522} 
\emlineto{23.409}{55.428} 
\emmoveto{23.409}{55.418} 
\emlineto{22.848}{55.315} 
\emmoveto{22.848}{55.305} 
\emlineto{22.277}{55.193} 
\emmoveto{22.277}{55.183} 
\emlineto{21.695}{55.059} 
\emmoveto{21.695}{55.049} 
\emlineto{21.102}{54.911} 
\emmoveto{21.102}{54.901} 
\emlineto{20.494}{54.748} 
\emmoveto{20.494}{54.738} 
\emlineto{19.870}{54.565} 
\emmoveto{19.870}{54.555} 
\emlineto{19.229}{54.358} 
\emmoveto{19.229}{54.348} 
\emlineto{18.566}{54.121} 
\emmoveto{18.566}{54.111} 
\emlineto{17.877}{53.844} 
\emmoveto{17.877}{53.834} 
\emlineto{17.157}{53.516} 
\emmoveto{17.157}{53.506} 
\emlineto{16.397}{53.112} 
\emmoveto{16.397}{53.102} 
\emlineto{15.585}{52.586} 
\emlineto{15.585}{52.596} 
\emmoveto{15.585}{52.586} 
\emlineto{14.698}{51.879} 
\emlineto{14.698}{51.889} 
\emmoveto{14.698}{51.879} 
\emlineto{13.699}{50.783} 
\emlineto{13.699}{50.793} 
\emmoveto{13.699}{50.783} 
\emlineto{12.521}{48.512} 
\emlineto{12.521}{48.522} 
\emmoveto{12.521}{48.512} 
\emlineto{12.000}{42.587} 
\emlineto{12.000}{42.597} 
\emmoveto{12.000}{42.587} 
\emlineto{12.637}{38.445} 
\emlineto{12.637}{38.455} 
\emmoveto{12.637}{38.445} 
\emlineto{13.301}{35.843} 
\emlineto{13.301}{35.853} 
\emmoveto{13.301}{35.843} 
\emlineto{13.918}{33.893} 
\emlineto{13.918}{33.903} 
\emmoveto{13.918}{33.893} 
\emlineto{14.491}{32.325} 
\emmoveto{14.491}{32.315} 
\emlineto{15.027}{30.995} 
\emmoveto{15.027}{30.985} 
\emlineto{15.531}{29.843} 
\emmoveto{15.531}{29.833} 
\emlineto{16.009}{28.828} 
\emmoveto{16.009}{28.818} 
\emlineto{16.464}{27.923} 
\emmoveto{16.464}{27.913} 
\emlineto{16.898}{27.109} 
\emmoveto{16.898}{27.099} 
\emlineto{17.313}{26.371} 
\emmoveto{17.313}{26.361} 
\emlineto{17.712}{25.698} 
\emmoveto{17.712}{25.688} 
\emlineto{18.096}{25.082} 
\emmoveto{18.096}{25.072} 
\emlineto{18.466}{24.517} 
\emmoveto{18.466}{24.507} 
\emlineto{18.822}{23.996} 
\emmoveto{18.822}{23.986} 
\emlineto{19.167}{23.517} 
\emmoveto{19.167}{23.507} 
\emlineto{19.501}{23.073} 
\emmoveto{19.501}{23.063} 
\emlineto{19.824}{22.664} 
\emmoveto{19.824}{22.654} 
\emlineto{20.137}{22.285} 
\emmoveto{20.137}{22.275} 
\emlineto{20.440}{21.935} 
\emmoveto{20.440}{21.925} 
\emlineto{20.734}{21.611} 
\emmoveto{20.734}{21.601} 
\emlineto{21.020}{21.312} 
\emmoveto{21.020}{21.302} 
\emlineto{21.298}{21.035} 
\emmoveto{21.298}{21.025} 
\emlineto{21.568}{20.780} 
\emmoveto{21.568}{20.770} 
\emlineto{21.830}{20.546} 
\emmoveto{21.830}{20.536} 
\emlineto{22.085}{20.331} 
\emmoveto{22.085}{20.321} 
\emlineto{22.333}{20.134} 
\emmoveto{22.333}{20.124} 
\emlineto{22.574}{19.954} 
\emmoveto{22.574}{19.944} 
\emlineto{22.808}{19.791} 
\emmoveto{22.808}{19.781} 
\emlineto{23.036}{19.644} 
\emmoveto{23.036}{19.634} 
\emlineto{23.258}{19.511} 
\emmoveto{23.258}{19.501} 
\emlineto{23.474}{19.393} 
\emmoveto{23.474}{19.383} 
\emlineto{23.684}{19.289} 
\emmoveto{23.684}{19.279} 
\emlineto{23.889}{19.198} 
\emmoveto{23.889}{19.188} 
\emlineto{24.088}{19.119} 
\emmoveto{24.088}{19.109} 
\emlineto{24.281}{19.053} 
\emmoveto{24.281}{19.043} 
\emlineto{24.469}{19.001} 
\emmoveto{24.469}{18.991} 
\emlineto{24.652}{18.957} 
\emmoveto{24.652}{18.947} 
\emlineto{24.830}{18.926} 
\emmoveto{24.830}{18.916} 
\emlineto{25.002}{18.904} 
\emmoveto{25.002}{18.894} 
\emlineto{25.170}{18.891} 
\emmoveto{25.170}{18.881} 
\emlineto{25.333}{18.891} 
\emmoveto{25.333}{18.881} 
\emlineto{25.492}{18.900} 
\emmoveto{25.492}{18.890} 
\emlineto{25.645}{18.918} 
\emmoveto{25.645}{18.908} 
\emlineto{25.794}{18.948} 
\emmoveto{25.794}{18.938} 
\emlineto{25.938}{18.983} 
\emmoveto{25.938}{18.973} 
\emlineto{26.078}{19.027} 
\emmoveto{26.078}{19.017} 
\emlineto{26.214}{19.079} 
\emmoveto{26.214}{19.069} 
\emlineto{26.345}{19.141} 
\emmoveto{26.345}{19.131} 
\emlineto{26.472}{19.211} 
\emmoveto{26.472}{19.201} 
\emlineto{26.594}{19.289} 
\emmoveto{26.594}{19.279} 
\emlineto{26.713}{19.377} 
\emmoveto{26.713}{19.367} 
\emlineto{26.827}{19.469} 
\emmoveto{26.827}{19.459} 
\emlineto{26.937}{19.569} 
\emmoveto{26.937}{19.559} 
\emlineto{27.043}{19.679} 
\emmoveto{27.043}{19.669} 
\emlineto{27.144}{19.793} 
\emmoveto{27.144}{19.783} 
\emlineto{27.242}{19.915} 
\emmoveto{27.242}{19.905} 
\emlineto{27.336}{20.042} 
\emmoveto{27.336}{20.032} 
\emlineto{27.426}{20.178} 
\emmoveto{27.426}{20.168} 
\emlineto{27.511}{20.322} 
\emmoveto{27.511}{20.312} 
\emlineto{27.593}{20.471} 
\emmoveto{27.593}{20.461} 
\emlineto{27.671}{20.624} 
\emmoveto{27.671}{20.614} 
\emlineto{27.745}{20.786} 
\emmoveto{27.745}{20.776} 
\emlineto{27.816}{20.952} 
\emmoveto{27.816}{20.942} 
\emlineto{27.882}{21.127} 
\emmoveto{27.882}{21.117} 
\emlineto{27.944}{21.306} 
\emmoveto{27.944}{21.296} 
\emlineto{28.003}{21.494} 
\emmoveto{28.003}{21.484} 
\emlineto{28.058}{21.683} 
\emmoveto{28.058}{21.673} 
\emlineto{28.109}{21.879} 
\emmoveto{28.109}{21.869} 
\emlineto{28.156}{22.085} 
\emmoveto{28.156}{22.075} 
\emlineto{28.199}{22.291} 
\emmoveto{28.199}{22.281} 
\emlineto{28.239}{22.505} 
\emmoveto{28.239}{22.495} 
\emlineto{28.275}{22.728} 
\emmoveto{28.275}{22.718} 
\emlineto{28.306}{22.951} 
\emmoveto{28.306}{22.941} 
\emlineto{28.335}{23.183} 
\emmoveto{28.335}{23.173} 
\emlineto{28.359}{23.415} 
\emmoveto{28.359}{23.405} 
\emlineto{28.379}{23.656} 
\emmoveto{28.379}{23.646} 
\emlineto{28.396}{23.905} 
\emmoveto{28.396}{23.895} 
\emlineto{28.409}{24.154} 
\emmoveto{28.409}{24.144} 
\emlineto{28.418}{24.408} 
\emmoveto{28.418}{24.398} 
\emlineto{28.423}{24.671} 
\emmoveto{28.423}{24.661} 
\emlineto{28.424}{24.938} 
\emmoveto{28.424}{24.928} 
\emlineto{28.422}{25.209} 
\emmoveto{28.422}{25.199} 
\emlineto{28.415}{25.484} 
\emmoveto{28.415}{25.474} 
\emlineto{28.405}{25.764} 
\emmoveto{28.405}{25.754} 
\emlineto{28.390}{26.049} 
\emmoveto{28.390}{26.039} 
\emlineto{28.372}{26.342} 
\emmoveto{28.372}{26.332} 
\emlineto{28.349}{26.635} 
\emmoveto{28.349}{26.625} 
\emlineto{28.323}{26.933} 
\emmoveto{28.323}{26.923} 
\emlineto{28.293}{27.239} 
\emmoveto{28.293}{27.229} 
\emlineto{28.258}{27.549} 
\emmoveto{28.258}{27.539} 
\emlineto{28.220}{27.860} 
\emmoveto{28.220}{27.850} 
\emlineto{28.177}{28.179} 
\emmoveto{28.177}{28.169} 
\emlineto{28.130}{28.503} 
\emmoveto{28.130}{28.493} 
\emlineto{28.079}{28.827} 
\emmoveto{28.079}{28.817} 
\emlineto{28.023}{29.159} 
\emshow{24.980}{17.700}{} 
\emshow{1.000}{10.000}{-8.00e1} 
\emshow{1.000}{17.000}{-6.40e1} 
\emshow{1.000}{24.000}{-4.80e1} 
\emshow{1.000}{31.000}{-3.20e1} 
\emshow{1.000}{38.000}{-1.60e1} 
\emshow{1.000}{45.000}{0.00e0} 
\emshow{1.000}{52.000}{1.60e1} 
\emshow{1.000}{59.000}{3.20e1} 
\emshow{1.000}{66.000}{4.80e1} 
\emshow{1.000}{73.000}{6.40e1} 
\emshow{1.000}{80.000}{8.00e1} 
\emshow{12.000}{5.000}{-4.84e0} 
\emshow{23.800}{5.000}{9.56e1} 
\emshow{35.600}{5.000}{1.96e2} 
\emshow{47.400}{5.000}{2.97e2} 
\emshow{59.200}{5.000}{3.97e2} 
\emshow{71.000}{5.000}{4.98e2} 
\emshow{82.800}{5.000}{5.98e2} 
\emshow{94.600}{5.000}{6.99e2} 
\emshow{106.400}{5.000}{7.99e2} 
\emshow{118.200}{5.000}{9.00e2} 
\emshow{130.000}{5.000}{1.00e3} 
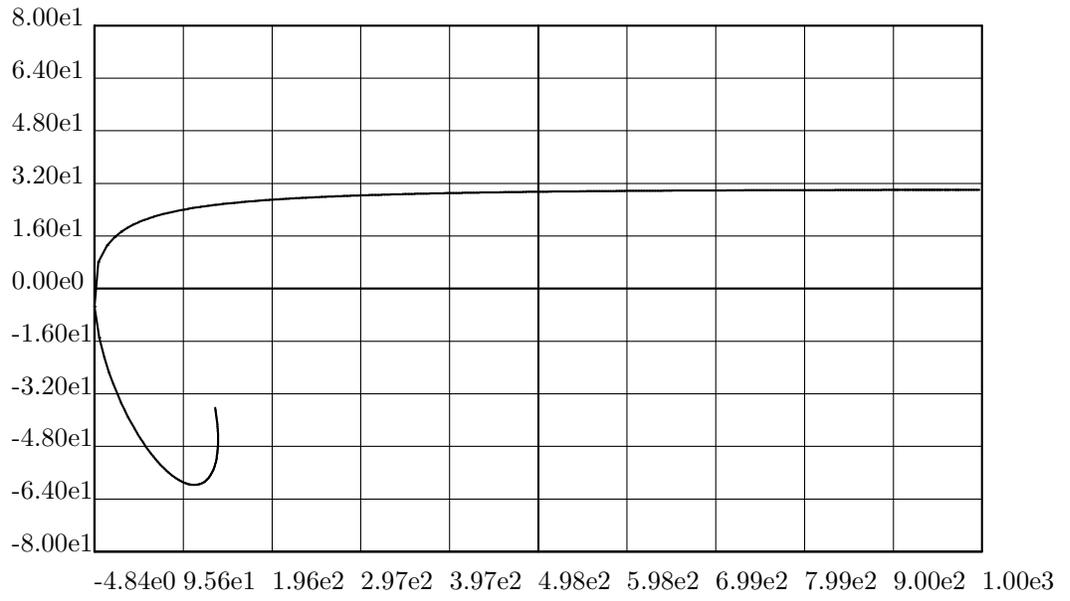
\centerline {\bf {Fig. 3}}

 \end{document}